\documentclass[a4paper,11pt]{article}
\pdfoutput=1 

\usepackage{jcappub} 
\usepackage{physics}
\usepackage[T1]{fontenc} 
\usepackage{aas_macros}
\usepackage{mathtools,  epsfig, lipsum}
\usepackage{lineno}

\def\beq{\begin{equation}}
\def\eeq{\end{equation}}
\def\bey{\begin{eqnarray}}
\def\eey{\end{eqnarray}}

\title{High-energy neutrinos from fallback accretion of binary neutron star merger remnants}

\author[a]{V. Decoene,}
\author[a,b,c]{C. Gu\'epin,}
\author[d]{K. Fang,}
\author[a]{K. Kotera,}
\author[e,f]{B. D. Metzger}

\affiliation[a]{Sorbonne Universit\'e, UPMC Univ. Paris 6 et CNRS, UMR 7095,  Institut d'Astrophysique de Paris, 98 bis bd Arago, 75014 Paris, France}
\affiliation[b]{Department of Astronomy, University of Maryland, College Park, MD  20742, USA}
\affiliation[c]{Joint Space-Science Institute, University of Maryland, College Park, MD  20742, USA}
\affiliation[d]{Einstein Fellow, Kavli Institute for Particle Astrophysics and Cosmology (KIPAC), Stanford University, Stanford, CA 94305, USA}
\affiliation[e]{Department of Physics and Columbia Astrophysics Laboratory, Columbia University,\\ Pupin Hall, New York, NY 10027, USA}
\affiliation[f]{Center for Computational Astrophysics, Flatiron Institute, 162 5th Ave, New York, NY 10010, USA}

\emailAdd{decoene@iap.fr}

\abstract{Following the coalescence of binary neutron stars, debris from the merger which remains marginally bound to the central compact remnant will fallback at late times, feeding a sustained accretion flow.  Unbound winds or a wide-angle jet from this radiatively-inefficient disk may collide with the comparatively slow dense kilonova ejecta released from an earlier phase.  Under the assumption that such interaction accelerate cosmic rays to ultra-high energies,  we numerically simulate their propagation and interactions through the dynamical ejecta. The hadronuclear and photo-hadronic processes experienced by particles produce isotropic high-energy neutrino fluxes, peaking at times $10^{3-4}\,$s, which we calculate for two sets of parameters. A first set is inspired by the observations of GW170817. In the second scenario, which we call optimistic, parameters are chosen so as to optimize the neutrino flux, within the range allowed by observation and theory.  We find that single sources can only be detected with IceCube-Gen2 for optimistic scenarios and if located within $\sim 4\,$Mpc. The cumulative flux could contribute to $\sim 0.5-10\%$  of the diffuse flux observed by the IceCube Observatory, depending on the fall-back power and the cosmic ray composition. The neutrino emission powered by fallback is nearly isotropic, and can be used for future correlation studies with gravitational wave signals.}

\begin{document}
\maketitle
\flushbottom

\section{Introduction}
The coincident detection of the binary neutron star (NS) merger GW170817 in gravitational waves and across the electromagnetic spectrum has launched a new era in multi-messenger astronomy \cite{2017PhRvL.119p1101A,2017ApJ...848L..13A, 2017ApJ...848L..12A}. The combined information provides new insights into the workings of particle acceleration and of the emissions taking place in compact objects and their environments.  Among the discoveries from GW170817 was thermal optical/infrared emission ("kilonova") powered by the radioactive decay of heavy nuclei synthesized in the merger ejecta (e.g.~\cite{Metzger+10}). 

One messenger whose future addition would further enrich our picture of neutron star merger events is the neutrino.  The environments of NS mergers are ideal {\it a priori} to produce copious neutrino fluxes, due to their large energy reservoirs, abundant source material to accelerate, and sizable radiative background fields for the accelerated particles to interact with.  Recent studies have shown that cosmic rays production up to energies near the so-called ``ankle'' feature in the Galactic cosmic ray spectrum ($\lesssim 10^{18}\,$eV) may be plausible in such merger events \cite{Kimura18,Rodrigues:2018tku}. 
Searches for GeV-EeV neutrinos directionally coincident with GW170817 within various time windows, were conducted with ANTARES, IceCube and the Pierre Auger Observatory, but no detections were reported \cite{2017ApJ...850L..35A}. The non-detection of GW170817 was found to be consistent with the models of neutrino production in binary NS mergers that had been postulated then \cite{Kimura17,2017ApJ...849..153F}.   

The end state of a binary neutron star merger is likely to be a black hole (BH) surrounded by a gaseous accretion disk, which powers a collimated relativistic jet producing a short gamma-ray burst (GRB) (e.g., \cite{1992ApJ...395L..83N, 2005A&A...436..273A, Rezzolla10, 2016ApJ...824L...6R,Liu2017}). Hence many of the existing models that estimate neutrino fluxes from NS mergers are related to GRBs observed at different viewing angles. The most promising neutrino-production mechanism from GRBs is related to the temporally extended X-ray/gamma-ray emission seen from a fraction of short GRBs (e.g.~\cite{Norris&Bonnell06}), for which the mild Lorentz factor of the outflows responsible for powering this emission enables a high meson production efficiency \cite{Kimura17, Kimura18, Biehl:2017qen, 2018PhRvD..98d3020K,Ahlers2019}. Other studies \cite{Gao:2013mcx,2017ApJ...849..153F} assume that the merger forms a magnetar capable of accelerating UHE particles, and calculate the abundant neutrino emission produced via interaction with the surrounding merger ejecta shell.

\begin{figure}
\centering
\includegraphics[width=1.0\columnwidth]{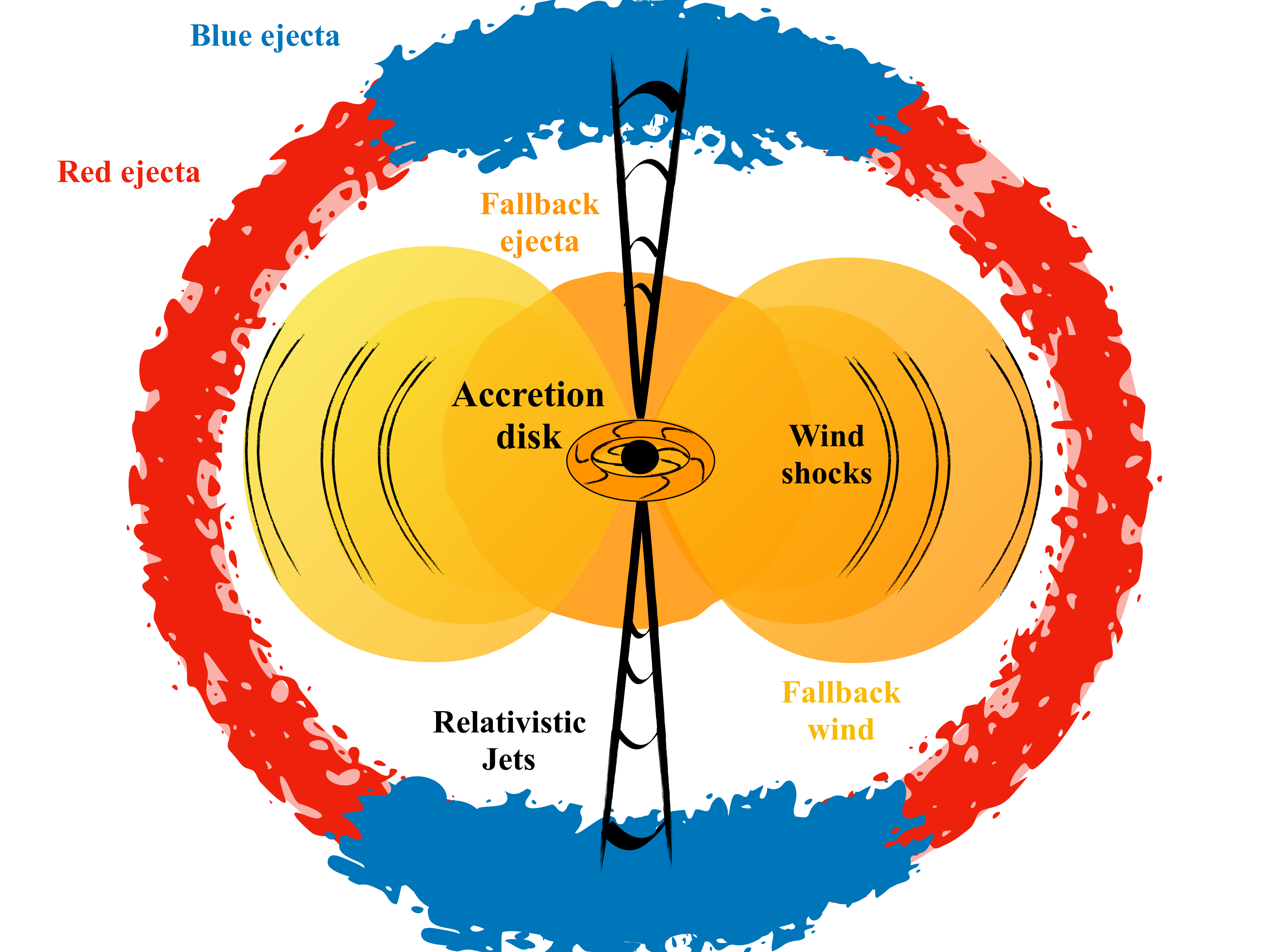}
\caption{Sketch of the regions of the neutron-star merger remnant at play for the acceleration and interaction of cosmic rays in our scenario. The red and blue envelopes indicate the location of the so-called blue and red kilonovae ejecta, that emit thermal UV/optical/IR radiation over timescales of hours to days (blue) and a week (red). Models related to the GRB jet have been explored in scenarios involving GRBs. In this work, we focus on the interaction of a fast wide-angle outflow from the accretion disk powered by late-time fall-back of merger debris, with the slowly-expanding red kilonova ejecta.  This interaction results in the dissipation of the accretion power as shocks or magnetic reconnection, accelerating relativistic particles, in a nebula behind the ejecta shell.}
\label{fig:sketch}
\end{figure}

In this work, we examine a different scenario for neutrino production from NS mergers.  At late times after the merger, a small fraction of the ejecta (e.g. from the tidal tail, or the promptly-formed accretion disk) remains marginally bound to the black hole, falling back to it over a range of timescale from seconds to days or longer.  Being too cold to cool effectively through (thermal) neutrino emission, yet too dense to cool through photon emission, the resulting accretion flow is radiatively-efficient (e.g.~\cite{DiMatteo+02}) and susceptible to powerful disk winds (e.g.~\cite{Rossi&Begelman09}) or a wide angle jet from the inner regions close to the black hole as depicted in Figure \ref{fig:sketch}.  These outflows will emerge into the cavity behind the higher mass ejecta shell released earlier during the merger and its immediate aftermath (that responsible for powering the observed kilonova emission) and will collide with its backside, generating shocks or forcing magnetic reconnection.  We assume that such an interaction will result in efficient cosmic-ray acceleration in a nebula behind the ejecta shell (see Section~\ref{section:acceleration}) and focus on the interaction of high-energy particles with the surrounding ejecta shell. We model for this purpose the evolution of the radiative and baryonic density of the dynamical ejecta over time, including heating via nuclear and fallback processes (Section~\ref{section:ejecta}). We calculate self-consistently via a Monte-Carlo code the energy losses and interactions experienced by cosmic rays, as well as their secondary neutrino fluxes as a function of time (Section~\ref{section:interactions}). We present our neutrino fluxes from single sources and integrated over whole populations for standard sets of source parameters, as narrowed down by the observation of GW170817, and for an optimistic scenario which can lead to an enhanced neutrino flux. We also estimate the rate of neutrino events expected from a merger population jointly detected in gravitational waves.

Quantities will be labelled $Q_x\equiv Q/10^x$ in cgs units unless specified otherwise, and except for particle energies which are in $E_x\equiv E/10^x\,$eV.

\section{Particle acceleration}\label{section:acceleration}
The merger of a NS-NS or NS-BH binary can lead to a spinning BH surrounded by an accretion disk of mass $0.01-0.1\,M_\odot$ (e.g., \cite{1997A&A...319..122R, 2003PhRvD..68h4020S, 2004MNRAS.351.1121R}). The accretion torus powers a collimated relativistic jet and creates a short gamma-ray burst (e.g., \cite{1992ApJ...395L..83N, 2005A&A...436..273A, Rezzolla10, 2016ApJ...824L...6R}). At early times ($\lesssim 0.1-1\,\rm s$) this disk can be effectively cooled by thermal neutrino emission \citep{1999ApJ...518..356P, 2001ApJ...557..949N,  0004-637X-858-1-52}. At later times, it becomes radiatively inefficient as the accretion rate drops \citep{0004-637X-657-1-383, 2008MNRAS.390..781M}. Some fraction ($\sim 10^{-3}-0.1\,M_\odot$) of the debris injected into the surrounding medium remains bound and gradually returns to the center as fallback matter. It accretes at a super-Eddington rate and can drive a powerful, radiation-driven wind \citep{2008MNRAS.390..781M}: we call this outflow the {\it late-time disk wind} (not to be confused with the early short-lived disk directly resulting from the coalescence). 

\subsection{Fallback mass and luminosity}\label{subsection:fallback_mass}
The ejecta which are marginally gravitationally bound to the black hole will return at late times $t \gg t_0$ at the following approximate rate (e.g., \cite{rosswog_2007,Metzger16,2017CQGra..34o4001F}):
\begin{align} \label{eq:Mfb}
\dot{M}_{\rm fb} &= \dot{M}_{{\rm fb}, t = t_{0}} \qty(\frac{t}{t_{0}})^{-5/3}\ ,
\end{align}
where
\begin{eqnarray}
\dot{M}_{{\rm fb,} t = t_{0}} &=& {M_{\rm fb}}\left[{\int_{t_{0}}^{t_{\rm end}} \qty(\frac{t}{t_{0}})^{-5/3} \dd{t} } \right]^{-1}
\\ &\sim& 3.3\times 10^{-1}  M_{\odot}\, {\rm s}^{-1} \ t_{0,-1}^{-1}\left[\left(\frac{t_{\rm 0, -1}}{t_{\rm end, 7}}\right)^{2/3}+1\right]^{-1} \qty(\frac{M_{\rm fb}}{0.05 M_{\odot}})\ ,\label{eq:Mdott0} 
\end{eqnarray}
Here $M_{\rm fb}$ is the total fall-back mass (normalized to a value similar to the total unbound ejecta mass inferred from the kilonova emission in GW170817), while $t_0$ and $t_{\rm end}$ correspond respectively to ad-hoc onset and fading times of the fallback process.  The kinetic luminosity of the outflow powered by fall-back accretion can then be parameterized as,
\begin{align} \label{eq:Lfb}
L_{\rm fb} &= \epsilon_{\rm fb}\,\dot{M}_{\rm fb}\,c^2 
\\ &\sim 1.3 \times 10^{46}\, {\rm erg\,s}^{-1}\, \epsilon_{\rm fb,-1}t_{3}^{-5/3} \,  \qty(\frac{M_{\rm fb}}{0.05\,M_{\odot}}) \ , \nonumber
\end{align}
where $\epsilon_{\rm fb} \sim 0.01-0.1$ is an efficiency factor (e.g., \cite{Metzger16}). The numerical values above, extrapolated to the first epoch at $t = 11$ hr of the observed kilonova from GW170817, leads to a fallback luminosity of $\sim 3\times 10^{43}\,{\rm erg}\,{\rm s}^{-1}$ for $\epsilon_{\rm fb} = 0.1$.  In Section~\ref{section:ejecta} we show how this luminosity heats the ejecta, leading to a thermal kilonova luminosity several orders of magnitude below the fallback luminosity. The parameters chosen above are thus compatible with the electromagnetic observations of GW170817 \cite{Margutti:2018xqd}, and in line with the recent numerical studies \cite{Siegel:2017nub}.

If the fall-back accretion-powered outflow expands faster than the slowest inner tail of kilonova ejecta ($\sim 0.1$ c in GW170817), then the disk outflows will catch up and shock with the inner edge of the ejecta shell, generating a nebula of gas behind it. Alternatively, if the flow is magnetically dominated, forced reconnection in the flow could lead to a similar dissipation of a portion of its Poynting flux. In this way, a portion of the fall-back power $L_{\rm fb}$ could be channelled into cosmic-ray acceleration and subsequent neutrino production. Here and in the following, we provide numerical quantities at $t=10^3\,$s, when neutrino production is most important in the case of pure proton injection for some of the ejecta parameter sets examined in this work. 

The amount of fallback mass is generally believed to be at most comparable to that of the unbound dynamical ejecta, in which case numerical simulations suggest an upper limit $M_{\rm fb}\lesssim 0.05 \, M_{\odot}$.  However, several factors could increase the amount of ejected mass  (and therefore the fallback mass). For instance, the dynamical ejecta is larger if the neutron stars are spinning rapidly at the time of merger (e.g. \cite{2019arXiv190605288E}) or for a particularly low binary mass ratio $q \ll 0.7$.
Although such properties are not compatible with the Galactic population of double neutron systems, such constraints may not apply to the extra-galactic population. Indeed, the second binary neutron star merger discovered by Advanced
LIGO, GW190425, possessed a total binary mass far in excess of known Galactic double neutron star systems, thus hinting at selection biases in the Galactic sample or wider diversity in the properties of the extragalactic binary population~\cite{2020arXiv200101761T}.  Numerical simulations of the merger itself also do not generally account for fallback of matter ejected from the post-merger accretion disk winds. The disk wind ejecta can be very large; hence $M_{\rm fb}\sim 0.05-0.1 \,M_\odot$ is not unreasonable, e.g. \cite{Siegel:2017nub}, and this likely produced most of the ejecta in GW170817. In light of these arguments, we will also examine in this study a case with a higher level of total fallback mass $M_{\rm fb}=0.1\,M_\odot$.

\subsection{Particle acceleration in the late-time accretion disk}
The late-time disk wind arises from an radiatively-inefficient accretion flow \cite{Narayan:1994et}, in which one may expect stochastic acceleration to take place (e.g., \cite{Lynn:2014dya}). On the other hand, accelerated particles can experience important energy losses on the dense baryonic background. The disk can be parametrized as a function of the black hole mass $M_{\rm BH}$, disk radius $R_{\rm disk}$, and Keplerian velocity $v_{\rm r}=\alpha_{\rm r} ({G M_{\rm BH}}/{R_{\rm disk}})^{1/2}$, where $\alpha_{\rm r}$ is the alpha parameter \cite{Shakura:1972te} and the accretion or fallback mass $M_{\rm fb} \sim 5 \times 10^{-2} M_{\odot}$ \cite{rosswog_2007}.
Assuming that the disk radius scales as the gravitational radius $R_{\rm disk} = r R_{\rm g}$ with $R_{\rm g} = GM_{\rm BH}/c^{2}$, one can express the baryonic density in the disk

\begin{align}
n_{\rm p, disk} &= \frac{\dot{M}_{\rm fb}}{2 \pi R_{\rm disk}^{2} v_{\rm r} m_{\rm p}} 
\\ &= 1.2 \times 10^{27} {\rm cm}^{-3} \ r_{1}^{-3/2} \alpha_{{\rm r},-1}^{-1} \ t_{3}^{-5/3} \qty(\frac{M_{\rm bh}}{6 M_{\odot}})^{-3/2} \qty(\frac{M_{\rm fb}}{0.05 M_{\odot}}) \ . \nonumber
\end{align}
Such high densities lead to drastically short hadronic interaction timescales for ultra-high energy cosmic rays: $t_{\rm pp}= (n_{\rm p, disk} \sigma_{\rm pp} c \kappa_{\rm pp})^{-1}  \sim 4.2 \times 10^{-12} \, {\rm s} \ n_{\rm p, disk, 27}^{-1} [\sigma_{\rm pp}\qty(1\,{\rm EeV})/\sigma_{\rm pp} ]$,
with the proton-proton interaction cross section $\sigma_{\rm pp} =  6.6 \times 10^{-26} \, {\rm cm}^2$ and inelasticity $\kappa_{\rm pp}\sim 0.5$ at $1\,$EeV~\cite{PhysRevD.74.034018}. These inevitable energy losses prevents any kind of acceleration process to succeed in the disk environment, even by invoking very efficient magnetic reconnection mechanisms in the magnetic rotational instability turbulence as was studied by \cite{2012ApJ...755...50R,Hoshino:2013pza,Hoshino:2015cka,kimura_2015}.\\

\subsection{Particle acceleration in the outer fallback wind region}\label{subsection:acc_outer}
A mildly-relativistic wind powered by fallback can be launched from the accretion disk and propagate out to large distances $r > 10^2- 10^3$. This region could be an alternative promising region for efficient acceleration. Indeed, the encounter of this outflow with the slower but higher mass outer ejecta would produce a shocked shell \cite{2013ApJ...772...30D}. Particles could in principle be shock-accelerated in this region. One may caution however that these internal shocks may be radiation-mediated due to the high optical depths at the early times considered here (see Section~\ref{section:ejecta}). Radiation-mediated shocks are not efficient to accelerate relativistic particles, unless a neutron-proton conversion mechanism is invoked \cite{Derishev:2003qi,Kashiyama:2013ata,Li:2015pao}.  On the other hand, the optical depth across the shocked region is likely to be much smaller than through the entire kilonova ejecta shell and thus the reverse shock which acts to decelerate the wind may still be collisionless, even at relatively early times.

Particle Acceleration could also take place via other processes in this region, e.g., 
via magnetic reconnection (see e.g., \cite{xiao2016} in the case of binary white dwarf mergers). It is not trivial to infer the level of large-scale magnetization in the post-merger phase of a binary neutron-star. Most numerical relativity simulations focus indeed on the disk formed after merger \cite{2016PhRvD..94f4012K,2017PhRvD..95f3016C,2019PhRvD.100b3005C}, following its evolution up to typically $\sim 100$\,ms. 
The magnetic field should be amplified by several orders of magnitude because of turbulence developing in the fluid during and after the merger (mainly via Kelvin-Helmholtz instability).  Although this is difficult to resolve numerically because of the small length-scales involved, the effects of such large magnetic fields have been studied for example in Refs.~\cite{2017PhRvD..95f3016C,2019PhRvD.100b3005C}. These works indicate that the equatorial outflow can be Poynting-flux dominated and hence enable reconnection processes \cite{comisso2019interplay,guo2015particle,Sironi_2014}.

For radii less than the diffusion radius, $R < R_D$, it is likely that any magnetic reconnection process is suppressed, due to the high photon drag. In this regime, radiation pressure works against the development of turbulence and against regions of approaching opposite magnetic polarity. Beyond $R_D$, however, radiation pressure starts to drop, and reconnection may start to occur, although the transition threshold from one regime to the other is not well-known (\cite{Uzdensky2011}, and references therein). The Thomson optical depth at the diffusion radius $\tau_{\rm T} (R_{\rm D}) \sim 1/\beta_{\rm wind}$ remains above unity for at least two orders of magnitude in radii beyond $R_{\rm D}$, and the dependence of the magnetic reconnection rate on the flow parameters in this still optically thick regime is speculative.

In the following, we will estimate the maximum achievable acceleration energy assuming that acceleration can operate. We will discuss how the numerical results would change depending on the considered acceleration mechanism.

The magnetic energy in the acceleration zone is assumed to be a fraction $\epsilon_B$ of the bulk kinetic energy of the wind dissipated into this region
\begin{align}\label{eq:Bsh_1}
\frac{B_{\rm acc}^2}{8 \pi} 4 \pi \, R_{\rm acc}^2 \, c =  \frac{1}{2}\epsilon_B \,  \epsilon_{\rm fb} \dot{M}_{\rm fb} (\beta_{\rm wind}c)^2\ ,
\end{align}
where the size of the region $R_{\rm acc} = r R_g$, with $r = 10^3$. The magnetic field strength in the region can then be expressed as
\begin{align} \label{eq:Bsh_2}
B_{\rm acc} &= \sqrt{\epsilon_{B} \epsilon_{\rm fb} \frac{\dot{M}_{\rm fb} \, \beta_{\rm wind}^2 c}{R_{\rm acc}^2}} 
\\ &= 1.5  \times 10^7 {\rm\, G} \ \epsilon_{B, -2}^{1/2} \epsilon_{\rm fb, -1}^{1/2} r_{3}^{-1} \beta_{\rm wind, -1} t_{3}^{-5/6} \qty(\frac{M_{\rm bh}}{6 M_{\odot}})^{-1} \qty(\frac{M_{\rm fb}}{0.05 M_{\odot}})^{1/2}\ . \nonumber
\end{align}
The acceleration timescale can be written as $t_{\rm acc} = gt_{\rm L}$, where $t_{\rm L}$ is the Larmor timescale and $g\ge 1$ the acceleration inefficiency factor. $g= 1$ corresponds to a maximally efficient acceleration process. As function of the particle energy $E$, the acceleration timescale then reads
\begin{align} \label{eq:t_acc}
t_{\rm acc} &= \frac{g\,E}{\beta_{\rm wind} c Z e B_{\rm acc}(t)} 
\\ &\sim 4.1 \times 10^{-1} {\rm\, s} \ E_{18} \, \beta_{\rm wind, -1}^{-2} \, Z^{-1} \ t_{3}^{5/6} \epsilon_{B, -2}^{-1/2} \epsilon_{\rm fb, -1}^{-1/2} r_{3} \qty(\frac{M_{\rm bh}}{6 M_{\odot}}) \qty(\frac{M_{\rm fb}}{0.05 M_{\odot}})^{-1/2}\ , \nonumber
\end{align} 
where $Z$ is the charge particle and $e$ the elementary charge. The above numerical estimate assumes that efficient acceleration operates with an inefficiency factor $g=1$. For shock acceleration for instance, this would correspond to a Bohm diffusion regime. Less efficient processes and other types of acceleration mechanisms could be considered by setting $g> 1$ \cite{2009JCAP...11..009L}.
For instance, for subrelativistic reconnection flows 
$g\gtrsim 10$ \cite{2010MNRAS.408L..46G}. 

One may estimate the baryonic density of the outflow as
\begin{align}
n_{\rm p, wind} &=  \frac{3\,\dot{M}_{\rm fb}}{4 \pi  (\beta_{\rm wind} \, c)^3\,t^2 \, m_{\rm p}} \\ 
&=  7.5 \times 10^{14}\, {\rm cm}^{-3} \ \beta_{\rm wind, -1}^{-3} \ t_{3}^{-14/3} \qty(\frac{M_{\rm bh}}{6 M_{\odot}})^{-2} \qty(\frac{M_{\rm fb}}{0.05 M_{\odot}}) \ . \nonumber
\end{align}
In the wind, at these times, the baryonic density is no longer an issue for particle acceleration since the interaction timescale for UHECR will be $t_{\rm pp} =  3.4\, {\rm s}\  \ n_{\rm p, 14.6}^{-1} [\sigma_{\rm pp}\qty(1\,{\rm EeV})/\sigma_{\rm pp} ]  > t_{\rm acc}$. On the other hand, the strong magnetic fields will also induce synchrotron cooling with timescale
\begin{align} \label{eq:t_sync}
t_{\rm sync} &= 6 \pi \qty(\frac{m_{\rm p}}{m_{\rm e}})^3 \frac{A^3}{Z^4} \frac{\qty(m_{\rm p} c^2)^2}{\sigma_{\rm T} c \beta B_{\rm sh}(t)^2 \ E} 
\\ &\sim 2.9 \times 10^{-3} \,{\rm s} \ E_{18}^{-1} \, A^{3} Z^{4} \ t_{3}^{5/3} \ , \nonumber
\end{align}
with $A$ the mass particle number, $\sigma_{\rm T}$ the Thomson cross section and $\beta$ the particle velocity ($\beta \to 1$ as ultra relativistic particle). Equation~\eqref{eq:t_sync} demonstrates that synchrotron cooling will be the main limiting factor to cosmic rays acceleration in this region. The maximal acceleration energy, computed for $t_{\rm acc}=t_{\rm syn}$, reads
\begin{align} \label{eq:Emax}
E_{\rm max} &= \sqrt{\frac{6 \pi \chi} {g(\chi^2 - 1)} \qty(\frac{m_{\rm p}}{m_{\rm e}})^3 \qty(\frac{A}{Z})^3 e} \frac{m_{\rm p} c^2}{\sqrt{\sigma_{\rm T} B_{\rm sh}}} 
\\ &\sim 1.2 \times 10^{17} {\rm\, eV} A^{3/2} Z^{-3/2} t^{5/12}_{3} \nonumber \ .
\end{align}
One notices that the rapid decrease of the magnetic field with time implies that particles can be accelerated up to increasingly higher energies at later times. 
It appears from this equation that the efficiency of the acceleration process will only influence the maximum energy by a factor of $1/\sqrt{g}$, with $g\gtrsim 10$ for reconnection/shear acceleration/non-Bohm diffusion. We assume in the following that $g=1$, although one should bear in mind that $E_{\rm max}$ might be overestimated.

As discussed in Ref.~\cite{Kimura18}, the diffusion timescale in the shock region can be estimated as $t_{\rm diff}\propto t_{\rm acc}$ in the Bohm regime. It is thus possible that diffusion prevents cosmic rays from escaping from the shock, except at the highest energies, naturally leading to hard or monoenergetic spectra peaking at $E_{\rm max}$. In the following, we will adopt a power-law spectrum $\propto E^{-\alpha}$ for the accelerated particles, with an index of either $\alpha=2.1$ (as commonly assumed for shock-acceleration) and $1.5$ (to depict a harder spectrum due to diffusion), with maximum energy $E_{\rm max}$. The hard spectral index also has the advantage of mimicking spectra obtained with other types of acceleration mechanisms that could be taking place, such as magnetic reconnection \cite{Zweibel09}. \\


\subsection{Particle acceleration in the corona}
In addition to fallback outflows, acceleration may happen in the corona of the accretion disk.
The corona is filled with a plasma coming from the accretion disk, this plasma is subject to turbulent phenomena leading to resonant Alfv\'en waves modes, that can accelerate particles stochastically. On the other hand, it is possible that the corona is immersed in an intense radiation field \cite{1998PASJ...50..325M}, leading to photo-hadronic losses that hinder efficient acceleration process.\\

\section{Radiative and baryonic backgrounds of the dynamical ejecta}\label{section:ejecta}
The neutron-star merger ejects unbound matter, that can be observed as a kilonova \cite{Metzger16}. On the polar axis, the lanthanide-free material induces light $r$-process nuclear reactions, that power a blue day-long emission.
The merger ejecta received sustained heating due to the radioactive decay of $r$-process nuclei, as well as due to outflows from the fallback accretion disk. Cosmic rays accelerated in the inner regions of the ejecta will experience synchrotron cooling and interactions on the radiation field and the baryonic material of this red ejecta. In order to evaluate the energy loss and interaction rates, we model the evolution of the thermal radiation background, magnetic field strength and baryon density of the merger ejecta. 

In the following, we compute our numerical values based on two fiducial parameter sets. One stems from the observation of GW170817, referred to as ``GW170817-like'', with total ejecta mass $M_{\rm ej}=10^{-2}$ and $\beta_{\rm ej}=0.3$. The other, referred to as ``optimistic'', is chosen to enhance the neutrino flux, with a lighter ejecta mass $M_{\rm ej}=10^{-4}$ and $\beta_{\rm ej}=0.3$ \cite{drout_light_2017,Shibata:2019wef,Villar:2017wcc}. The more dilute radiative and baryonic environments induce less secondary meson cascades, which inhibit efficient neutrino production. References~\cite{Metzger16,Shibata:2019wef} review the mass and velocity ranges allowed for the various binary-neutron-star systems studied analytically and numerically. Following the discussion in Section~\ref{subsection:fallback_mass}, the fallback mass is set to $M_{\rm fb}=0.05\,M_\odot$, and to $M_{\rm fb}=0.1\,M_\odot$, for the GW170817-like and optimistic scenarios respectively.
Table~\ref{table:KN_parameters} summarizes the key parameters chosen for these two scenarios.

\begin{table}[tbp]
\center
\begin{tabular}{lccccccc}
\hline
   Scenario & $M_{\rm ej}$ & $\beta_{\rm ej}$ & $M_{\rm fb}$ & $\beta_{\rm wind}$ & $\epsilon_{\rm fb}$ &  $\dot{n}_0$ & ${\cal R}(z)$\\
   &$[M_{\odot}]$& &$[M_{\odot}]$& && [Gpc$^{-3}$yr$^{-1}$] & \\
    \hline
    GW170817-like & $\rm 10^{-2}$ & $0.3$ & $5 \times 10^{-2}$ & 0.1 & 0.1 & 600 &1\\
     Optimistic & $\rm 10^{-4}$ & $0.3$ & $1 \times 10^{-1}$ & 0.1 & 0.1 & 3000 & ${\cal R}_{\rm SFR}(z)$ \\
     \hline
\end{tabular}
\caption{Key parameters of binary neutron star merger remnants chosen for the two fiducial scenarios examined in this work: dynamical ejecta mass $M_{\rm ej}$ and velocity $\beta_{\rm ej}$, wind fallback mass $M_{\rm fb}$ and velocity $\beta_{\rm wind}$, population rate $\dot{n}_0$ and emissivity evolution with redshift ${\cal R}(z)$. GW170817-like model based on parameters obtained from the observation of GW170817. Optimistic case chosen so as to enhance the neutrino flux, while staying consistent with the allowed ranges for extragalactic populations of mergers. See Sections~\ref{subsection:fallback_mass} and \ref{section:ejecta}.}
\label{table:KN_parameters}
\end{table}

\subsection{Thermal photon background}
The ejecta material being in free expansion, its radius $R_{\rm ej}$ evolves linearly in time: $R=\beta_{\rm ej}c \,t$. The thermal energy $E$ of the ejecta evolves according to 
\begin{align} \label{eq:dEvdt}
\dv{E}{t} = - \frac{E}{R_{\rm ej}} \dv{R_{\rm ej}}{t} - \frac{E}{t_{\rm d}} +  \dot{Q}_{\rm r} + L_{\rm fb} \ .
\end{align}
The first term on the right hand side of Eq.\eqref{eq:dEvdt} describes the $P{\rm d}V\sim (E/V){\rm d}V$ work done by pushing the ejecta. The second term corresponds to the escape of thermal radiation from the mass layer, which can be written as
\begin{align} \label{eq:tescv}
t_{\rm d} &\approx \qty(\tau + 1) \frac{R_{\rm ej}}{c} = \qty( \frac{3 M_{\rm ej} \kappa}{4\pi R_{\rm ej}^{2}} + 1) \frac{R_{\rm ej}}{c} = \qty[\qty(\frac{t_{\rm d, 0}}{t})^2 + 1]\beta_{\rm ej} \, t \ ,
\end{align}
where we define the transparency time
\begin{align} \label{eq:tesc_0}
t_{\rm d, 0} &\equiv {\qty(\frac{3 M_{\rm ej} \kappa}{4 \pi (\beta_{\rm ej}c)^{2}})^{1/2}} 
\\ &\sim 0.89 \ { \rm days} \ \qty(\frac{M_{\rm ej}}{10^{-4} M_{\rm \odot}})^{1/2} \qty(\frac{\kappa}{10 \, {\rm cm}^{2} \ {\rm g}^{-1}})^{1/2} \ . \nonumber
\end{align}
Here we adopt $\kappa = 10$ g$^{-1}$ cm$^{2}$ as the opacity of the lanthanide-rich ejecta \cite{Villar:2017wcc}. 

The third and fourth terms in equation~(\ref{eq:dEvdt}) are the source terms of heating due to radioactivity $\dot{Q}_{\rm r}$ and fallback accretion $L_{\rm fb}$ respectively. We already described and expressed the fallback luminosity $L_{\rm fb}$ in section~\ref{section:acceleration} (Eq.~\ref{eq:Lfb}).

The heating from radioactive decay of heavy nuclei synthesised in the ejecta by $r$-process can be parametrised as \cite{barnes_radioactivity_2016}:
\begin{align} \label{eq:Qr_1}
\dot{Q}_{\rm r} &= M_{\rm ej} X_{\rm r} \dot{e}_{\rm r}\qty(t)\ ,
\end{align}
where $X_{\rm r}$ is the mass fraction of lanthanides in the ejecta and $\dot{e}_{\rm r}$ is the nuclear mass energy, which can be expressed as \cite{Korobkin2012}

\begin{align}  \label{eq:Qr_2}
\dot{e}_{\rm r}\qty(t) = 4\times10^{18} \epsilon_{\rm th} \left[0.5 - \frac{1}{\pi}\,\arctan{\qty(\frac{t - t_{0}}{\sigma})}\right]^{1.3} \textrm{\ erg\,s}^{-1}\textrm{\,g}^{-1}\ ,
\end{align}
with $t_{0} = 1.3$\,s and $\sigma  = 0.11$\,s, the starting time and the characteristic time of the decay, it can be noticed that the decay start only after $\sim 1$\,s due to the absorption of the free neutrons in the ejecta, and $\epsilon_{\rm th}$ the thermal efficiency of the nuclear processes, accounting for the energy deposit in the ejecta of the radioactive decay products of non-thermal beta-particles, alpha-particles, fission fragments, and gamma-rays by \cite{barnes_radioactivity_2016} and parametrised as follow

\begin{align}  \label{eq:Qr_3}
\epsilon_{\rm th} = 0.36 \, \qty[\exp\qty(-a \, t_{\rm day}) + \frac{\ln{\qty(1 + 2 b \, t_{\rm day}^{d})}}{2b \, t_{\rm day}^{d}}] \ ,
\end{align}
with $a = 2.19$, $b = 0.31$, $d = 1.52$. The numbers are valid for $M_{\rm ej} = 0.01M_{\odot}$ and initial velocity $v=0.3c$ (from \cite{barnes_radioactivity_2016}).

\begin{figure}
\center
\includegraphics[width=0.48\linewidth]{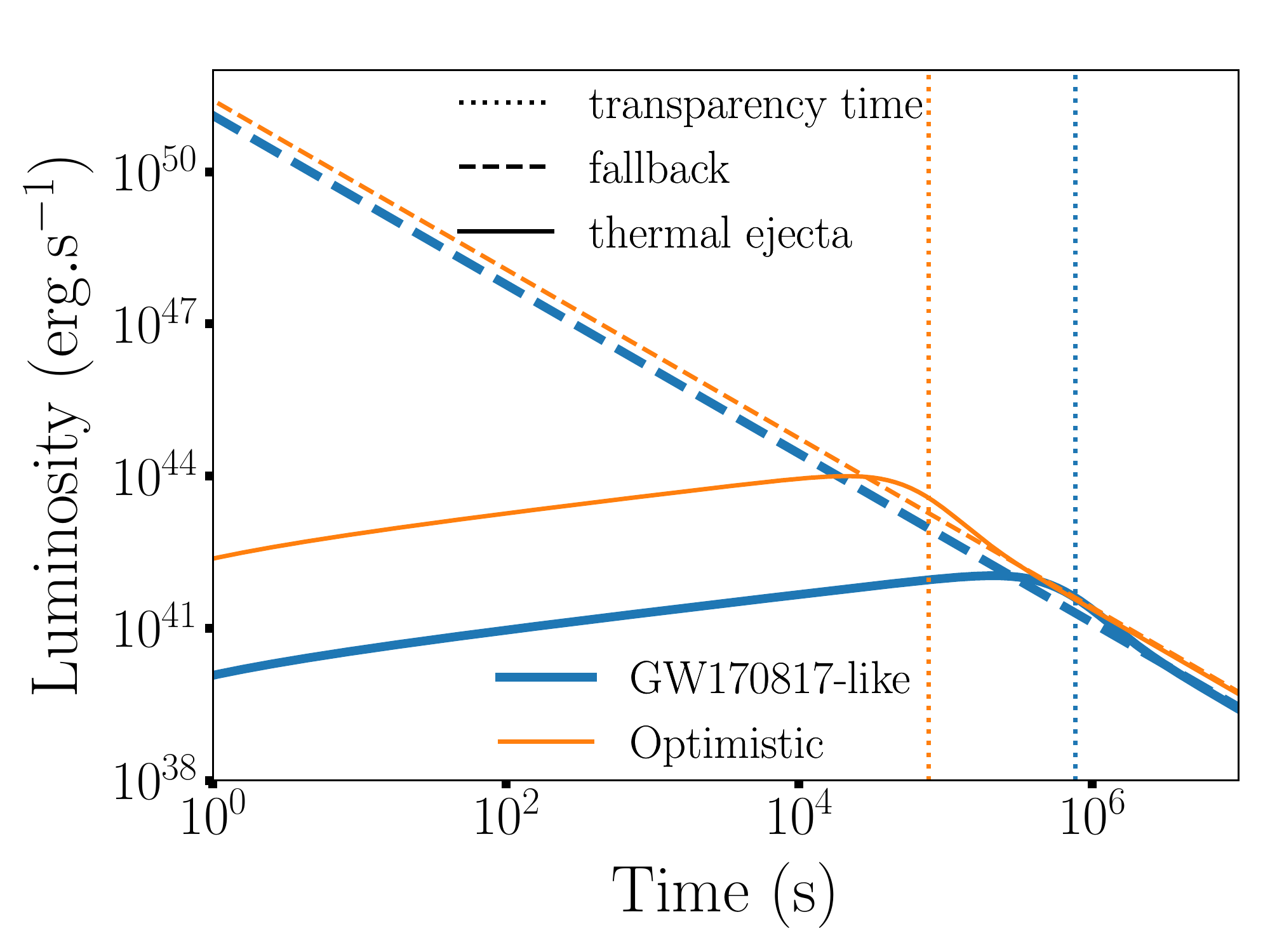}
\includegraphics[width=0.48\linewidth]{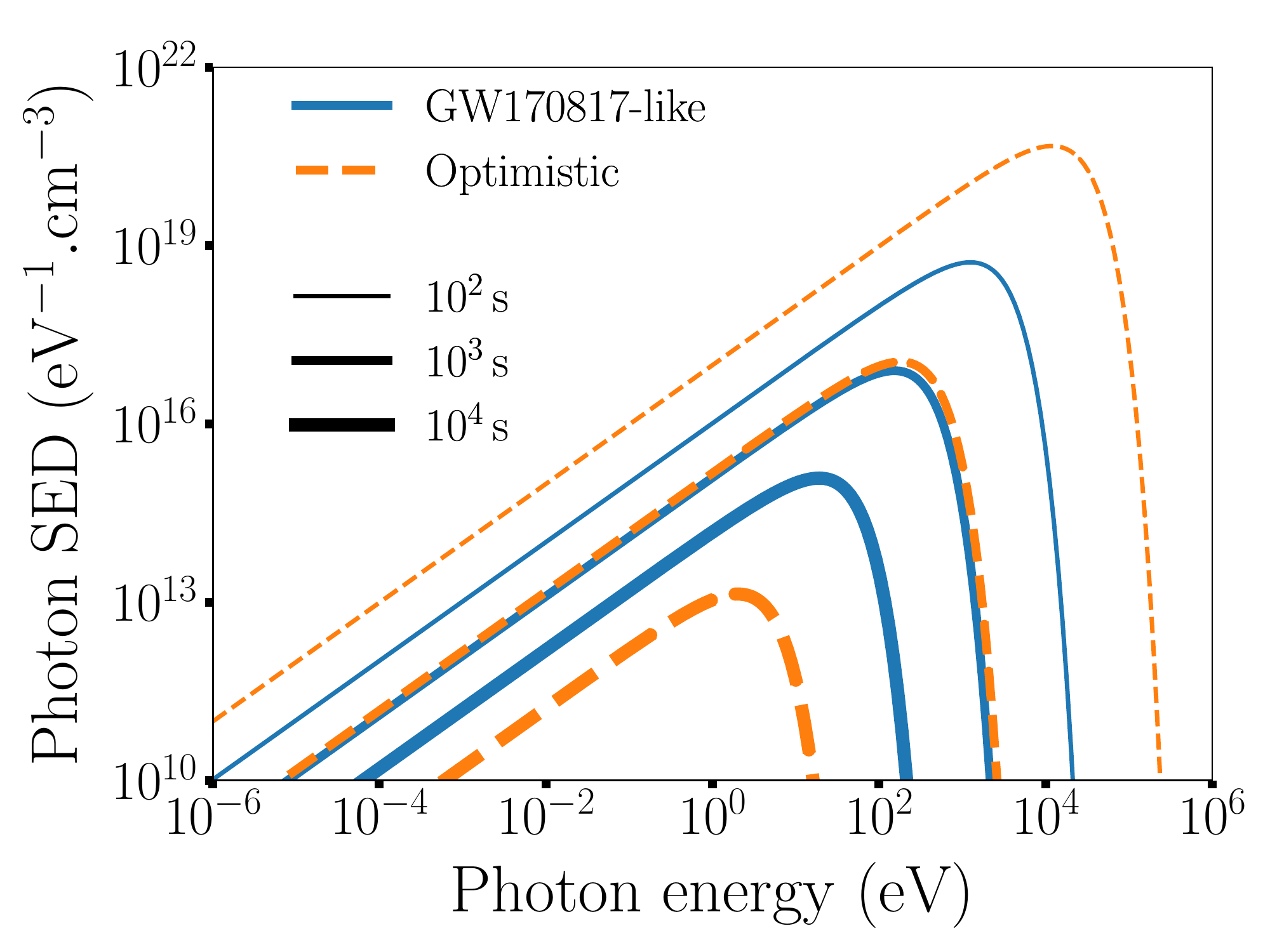}
\caption{Ejecta luminosity and spectral evolution in time for our two fiducial scenarios (Table~\ref{table:KN_parameters}). {\it Left:} Time evolution of the ejecta luminosity (solid), fallback luminosity (dashed). The vertical lines correspond to the transparency times $t_{\rm d,0}$, when the dynamical ejecta turns optically transparent. {\it Right:} Corresponding black body photon spectral energy density of the dynamical ejecta at various times.}
\label{fig:lum_thermalspec}
\end{figure}

The thermal luminosity is then given by

\begin{align} \label{eq:thermal_lum}
L_{\rm th} &= \frac{E}{t_{\rm d}}
\end{align}
and is shown in the left panel of Figure~\ref{fig:lum_thermalspec}. The luminosity first increases rapidly in time and reaches a regime where $L_{\rm th}\propto t^{1/3}$ for $t\ll t_{\rm d,0}$.
Considering a homogeneous repartition of nuclear reactions, one can assume that the ejecta emission follows a black-body, with a flux density at photon frequency $\nu$

\begin{align}\label{eq:Fnu}
F_\nu(t) = \frac{2\pi h \nu^3}{c^2}\frac{R_{\rm ej}}{D^2}\left( e^{h\nu/kT_{\rm eff}}-1\right) \ ,
\end{align}
where $D$ is the distance to the source and the effective temperature is computed following the Stefan-Boltzman law:

\begin{align}\label{eq:Teff}
T_{\rm eff} &= \qty(\frac{3E}{4 \pi a R_{\rm ej}^{3}})^{1/4}\ .
\end{align}
In principle, the radius of the shell $R_{\rm ej}$ in the equations above (Eqs.~\ref{eq:Fnu} and \ref{eq:Teff}) corresponds to the photosphere radius. We consider here that the photosphere radius can be approximated as the ejecta radius $R_{\rm ej}=\beta_{\rm ej}c \, t$. The thermal radiation spectra of the dynamical ejecta is presented at various times in the right panel of Fig.~\ref{fig:lum_thermalspec}. 

\subsection{Magnetic field strength}
The magnetic field strength can be computed following the same reasoning as in Eq.\eqref{eq:Bsh_1}, assuming that the magnetic energy density in the ejecta is sourced by the non-thermal inflow of the fallback in a volume of radius $R_{\rm ej}$:
\begin{align} \label{eq:Bej_1}
\frac{B_{\rm ej}^2}{8 \pi} \qty( 4 \pi \, R_{\rm ej}^2 \, c ) = \epsilon_{\rm B} L_{\rm fb} \ ,
\end{align}
leading to 
\begin{align} \label{eq:Bej_2}
B_{\rm ej} &= \sqrt{2 \epsilon_{\rm B} \epsilon_{\rm fb} \frac{\dot{M}_{\rm fb} \, c}{R_{\rm ej}^2}} 
\\ &\sim 3.3 \times 10^{3} \ {\rm G} \ \epsilon_{\rm B, -3}^{1/2} \epsilon_{\rm fb, -1}^{1/2} \qty(\frac{\beta_{\rm ej}}{0.3}) \ t_{3}^{-5/6} \qty(\frac{M_{\rm bh}}{6 M_{\odot}})^{-1} \qty(\frac{M_{\rm fb}}{0.05 M_{\odot}})^{1/2}\ . \nonumber
\end{align}
This magnetic field will determine the level of cooling processes inside the kilonova.

\subsection{Baryonic density}
The baryonic density determines the hadronic processes happening inside the kilonova ejecta. We compute the total mass of the ejecta in the volume of the ejecta, assuming that the mean mass number of nuclei in the ejecta is $\langle A\rangle\sim 100$. 
Indeed, r-process simulations driven by the observation of GW170817 indicate that nucleosynthesis is efficient, and most nuclei lie in the mass range $80\lesssim A \lesssim 200$ in the inner parts of the ejecta. The lanthanide mass fraction is smaller, within the range $X_{\rm r} \approx 10^{-3}-10^{-2}$ \cite{Kasen:2017sxr,Chornock:2017sdf}. 
The baryonic density then reads
\begin{align} \label{eq:density}
n_{\rm ej} &= \frac{3 M_{\rm ej}}{4\pi R_{\rm ej}^{3}} \frac{1}{\langle A\rangle m_{\rm p}}
\\ &\sim 6.5 \times 10^{13} \ {\rm cm}^{-3} \ \qty(\frac{\beta_{\rm ej}}{0.3})^{-3} \, t_{3}^{-3} \, \qty(\frac{M_{\rm ej}}{10^{-4} \, M_{\odot}}) \ . \nonumber
\end{align}

Studies also show that the decay of remnant free neutrons in the outer ejecta, where the velocities are important, produces a high abundance of Hydrogen, which dominates the composition in the outer layers of  ejecta~\cite{2016ApJ...829..110B}. This can have a mild effect on the hadronic interaction timescale, as the relevant quantity for its calculation is the column density, and the outer layers are important, given their large radii.

\section{Cosmic-ray interactions in the dynamical ejecta}\label{section:interactions}
Particles accelerated in the inner regions propagate in the equatorial ejecta (the so-called red kilonova), and experience various interaction and cooling processes, that are described in this section. In particular, they interact with the radiative and hadronic backgrounds presented in section~\ref{section:ejecta}. We calculate the secondary neutrino flux produced by the relevant interactions: i) analytically for protons as a basis for comparison, and ii) numerically using a Monte-Carlo propagation and interaction code. 

\subsection{Numerical setup}
We use the numerical propagation and interaction code developed in \cite{Guepin18b}, with modules from CRPropa3 \citep{alvesbatista16} and the code described in \cite{kotera09}, which accounts for all relevant interaction and energy-loss processes for nucleons and heavier nuclei. Nucleons experience pion production through photohadronic and hadronic interactions, as well as neutron and unstable nuclei decay. All charged particles, including charged pions and muons, undergo synchrotron, inverse Compton and Bethe-Heitler processes. The interaction cross sections and products are obtained from analytic formulae \citep[e.g.][]{r96, dermer09} or tabulated from \textsc{Sophia} \citep{Mucke00} for photopion production, \textsc{Talys} \citep{talys} for photonuclear interactions, and \textsc{Epos} \citep{WLP06} for  hadronic interactions. We assume that the photofragmentation products are similar to the products of hadronic interactions, which is reasonable to first approximation, as argued in \cite{Guepin18b}. In the following, we outline the main interactions, and provide analytical estimates of the proton mean free paths.

\subsection{Primary cosmic-ray interactions} \label{subsection:cosmic_ray_interactions}

\begin{figure}[tb]
\centering
\includegraphics[width=0.48\linewidth]{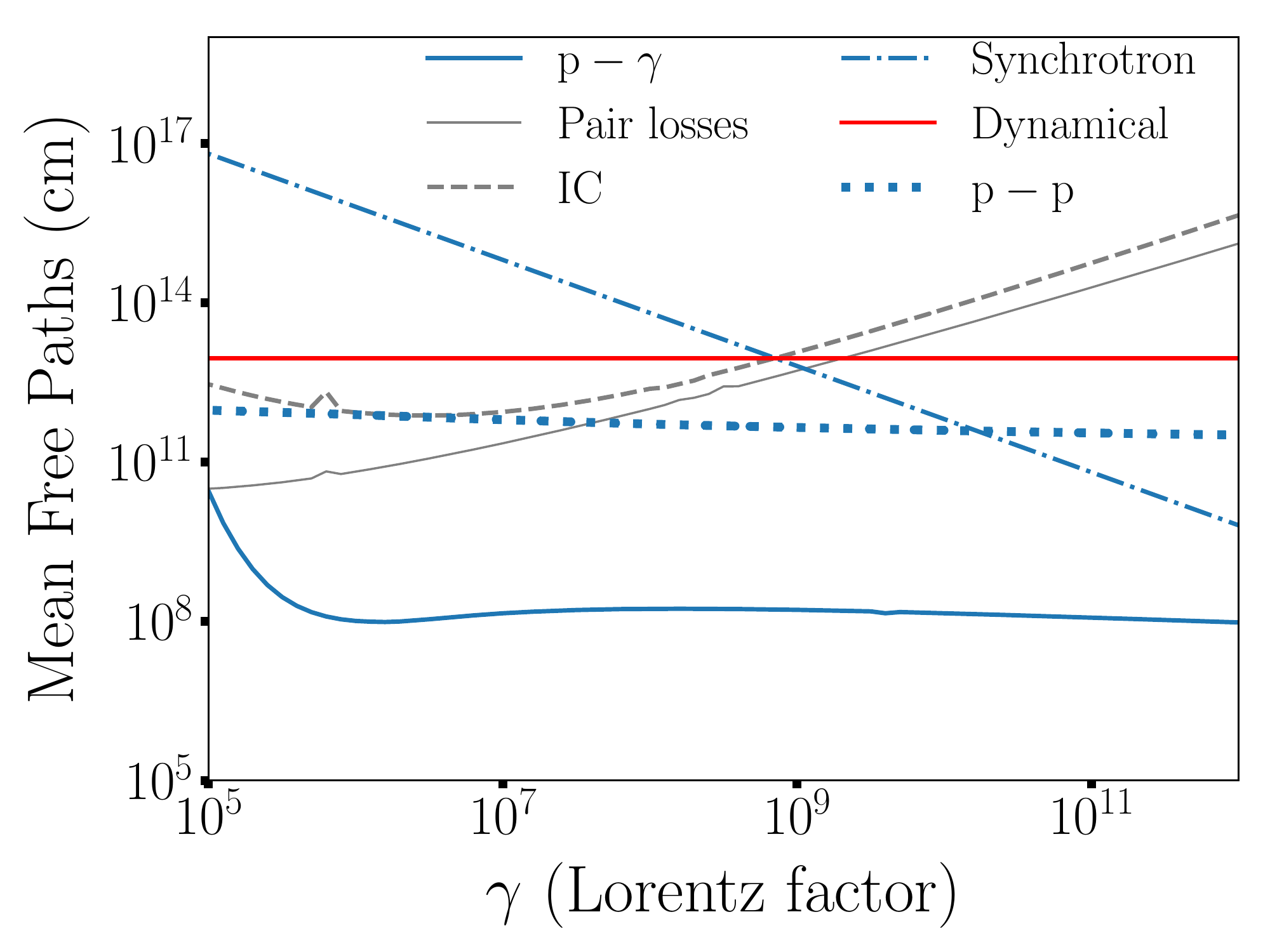}
\includegraphics[width=0.48\linewidth]{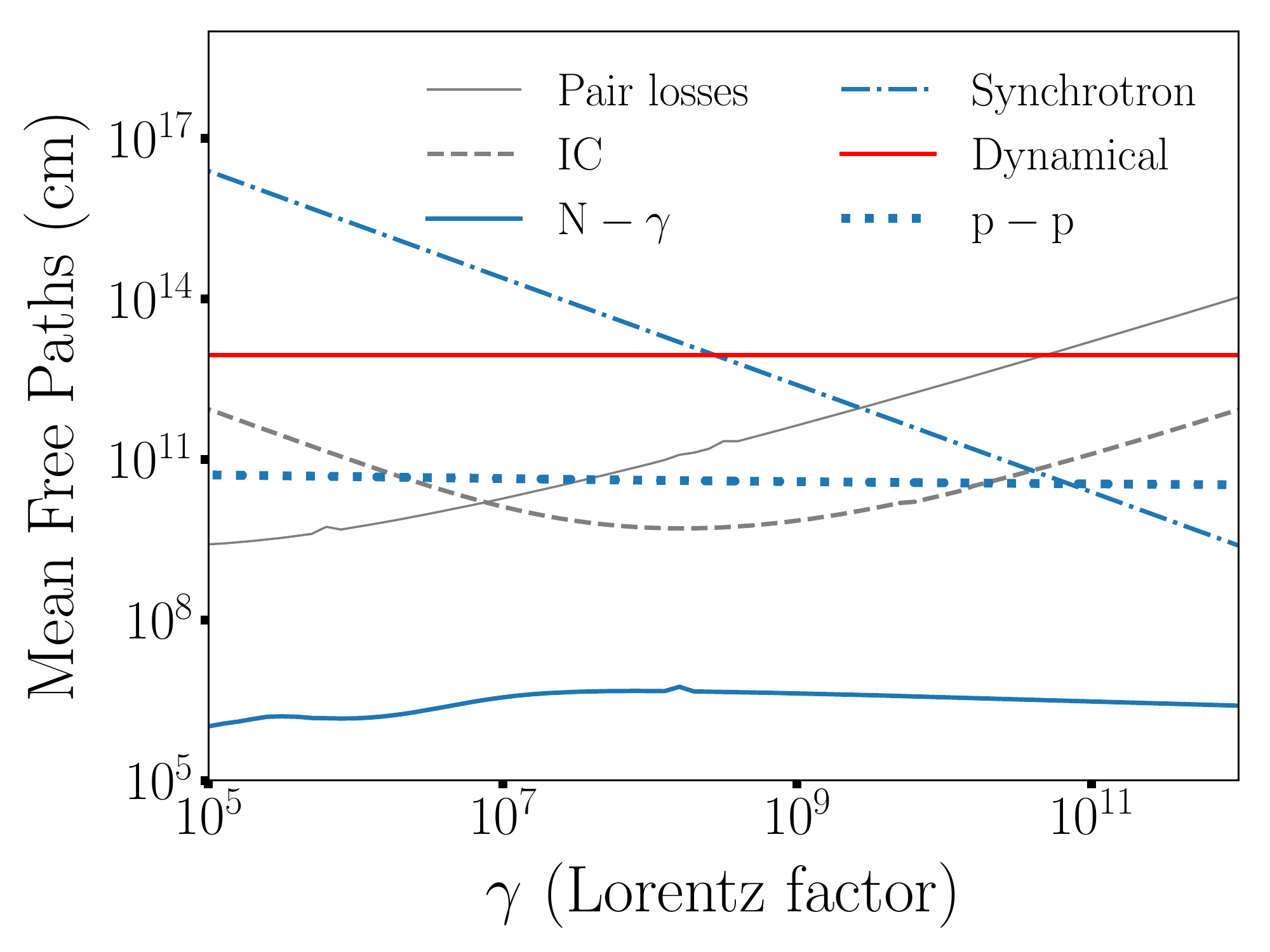}
\includegraphics[width=0.48\linewidth]{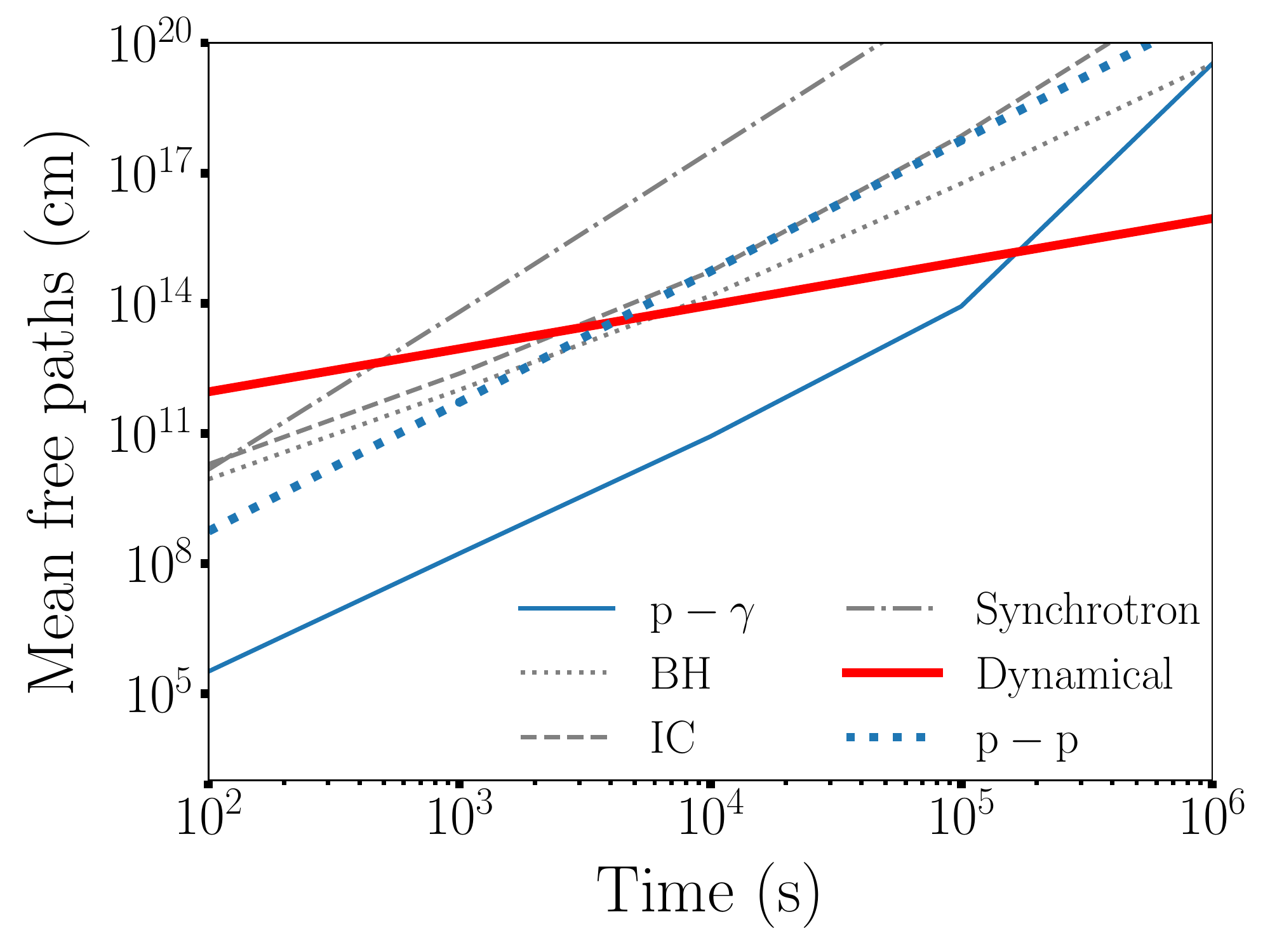}
\includegraphics[width=0.48\linewidth]{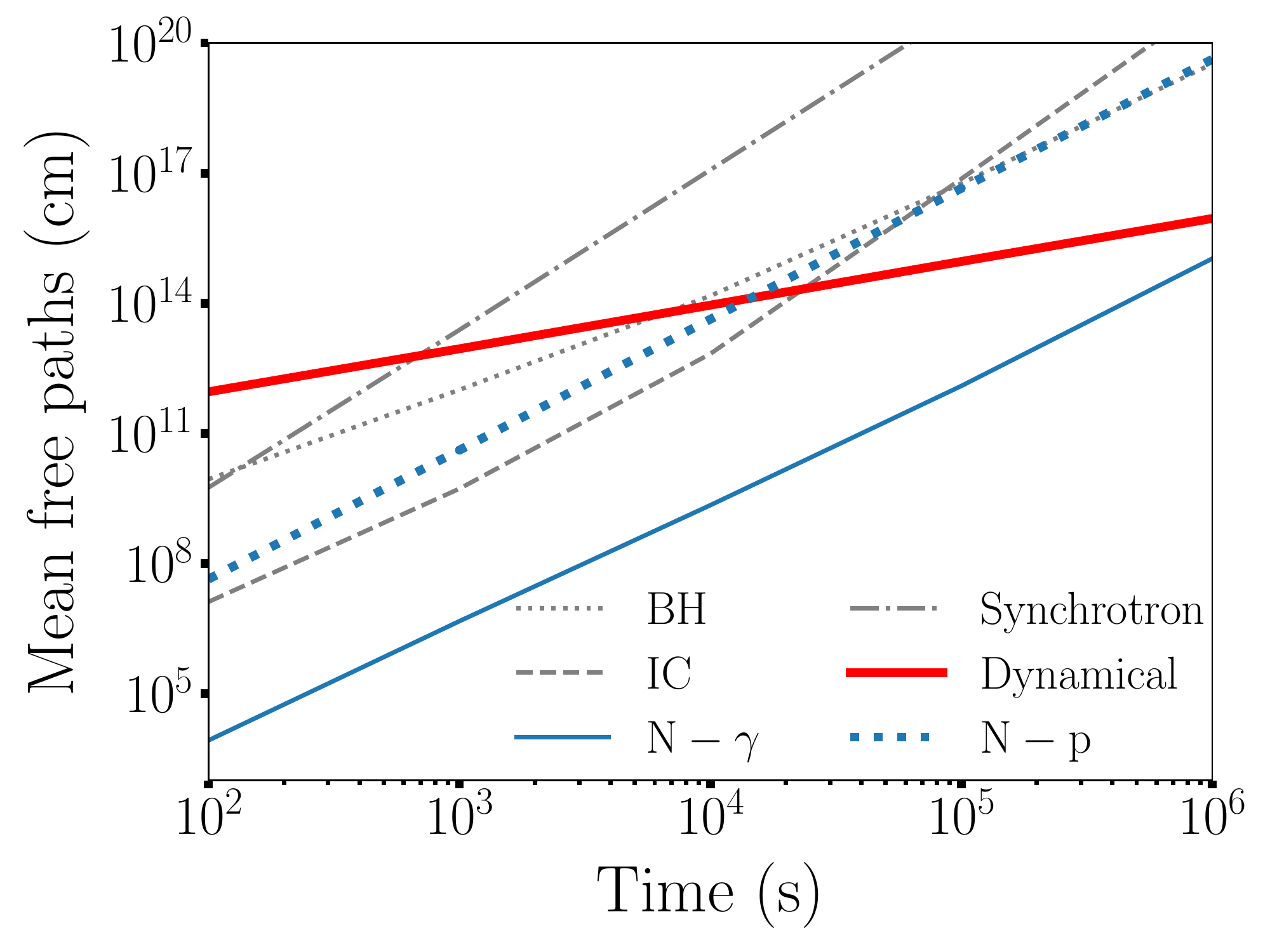}
\caption{Mean free paths for protons ({\it left}) and iron nuclei ({\it right}), for photohadronic and purely hadronic interactions, compared to the typical cooling lengths by inverse Compton, synchrotron, pair-production and dynamical expansion (optimistic scenario). {\it Upper panels:} Mean free paths as a function of particle Lorentz factor for $t=10^3$\,s. {\it Lower panel:} Mean free paths as a function of time for $\gamma = 10^8$.}
\label{fig:mfps}
\end{figure}

In the equatorial ejecta, the interaction with the radiative background leads to photomeson production, photodisintegration, inverse Compton (IC) and Bethe-Heitler processes. Charged particles undergo synchrotron energy losses, due to the magnetic field produced by the non-thermal heating from the inner regions. As shown in Eq.~\eqref{eq:Bej_2}, the intensity of this interactions is linked to the fallback luminosity $L_{\rm fb}$ and the ejecta expansion $R_{\rm ej}$. Moreover, the interaction with the baryonic background leads to purely hadronic interactions.

As shown in section~\ref{section:ejecta}, the equatorial ejecta is characterised by two main parameters: $M_{\rm ej}$ and $\beta_{\rm ej}$. The radiative and hadronic backgrounds depend on these two parameters, and so do the photonuclear and hadronic interactions. Therefore, $M_{\rm ej}$ and $\beta_{\rm ej}$ eventually influence the high-energy neutrino production, as photonuclear and hadronic interactions produce neutrinos, for instance through the channels ${\rm p} + \gamma \to \pi^{+} + {\rm n}$, (${\rm n} + \gamma \to \pi^{-} + {\rm p}$) and ${\rm p} + {\rm N} \to n_{\pi} \pi^{\pm} + n_{\rm p}{\rm p} + n_{\rm n}{\rm n}+ {\rm N'}$, and the decays $\pi^{\pm} \to \mu^{\pm} + \nu_{\mu}$ and $\mu^{\pm} \to e^{\pm} + \nu_{e} + \nu_{\mu}$.

In the following, we describe the energy-loss timescales $t_\star$, the mean free paths $l_\star = c t_\star$ and the interaction depths $\tau_\star = t_{\rm dyn}/t_\star$, where the subscript $\star$ corresponds to any interaction and $t_{\rm dyn} = R_{\rm ej}(t) / c$ characterises the expansion of the ejecta. These quantities allow us to identify dominant interaction processes and evaluate their efficiencies. In particular, an interaction depth larger than $1$ indicates an efficient interaction process. The photopion production energy-loss timescale $t_{\rm p \gamma}$ is
\begin{align} \label{eq:t_pgamma}
t_{\rm p \gamma}^{-1}\qty(\gamma_{\rm p}) = \frac{c}{2 \gamma_{\rm p}^2} \int_{0}^{\infty} \dd{\epsilon} \sigma_{\rm p \gamma}\qty(\epsilon) \kappa_{\rm p \gamma}\qty(\epsilon) \epsilon \int_{\epsilon/2 \gamma_{\rm p}}^{\infty} \dd{\bar{\epsilon}} \bar{\epsilon}^{-2} n_{\rm BB}\qty(\bar{\epsilon}) \, ,
\end{align}
where $\gamma_{\rm p}$ is the Lorentz factor of the accelerated proton, $\sigma_{\rm p \gamma}\qty(\epsilon)$ the photopion production cross section as function of the photon energy $\epsilon$, $\kappa_{\rm p \gamma}\qty(\epsilon)$ the proton inelasticity and $n_{\rm BB}\qty(\epsilon)$ the black body spectral energy density of the ejecta derived from Eq.\eqref{eq:Fnu}.

Using Eq.~\eqref{eq:t_pgamma} and assuming a Heaviside function for the cross section with pion production threshold $\epsilon_{\rm thres} = 145$\,MeV (see~\cite{Waxman:1997ti} Eq.~3), we derive an analytical estimate of the interaction depth of photopion production
\begin{align} \label{eq:pgamma_depth}
    \tau_{\rm p \gamma}\qty(\gamma_{\rm p}) &= \frac{8 \pi}{\qty(h c)^3} \sigma_{\rm p \gamma}\qty(\gamma_{\rm p})  \kappa_{\rm p\gamma} R_{\rm ej} \qty(k_{\rm B} T_{\rm ej})^3 \times \mathcal{I}\qty(\gamma_{\rm p} ; T_{\rm ej}) \, ,
    \\ &\sim 1.1 \times 10^4 \ \qty(\frac{\beta_{\rm ej}}{0.3}) \, t_{3} \, T_{\rm ej, 6}^{3}\qty(M_{\rm ej}, \beta_{\rm ej}) \, \left[\frac{\sigma_{\rm p \gamma}(\gamma_{\rm p}) {\kappa_{\rm p\gamma}}}{70 \, \mu {\rm barn}}\right] \, , \nonumber
\end{align}
where $h$ is the Planck constant, $T_{\rm ej}$ is given by Eq.\eqref{eq:Teff} and $\mathcal{I}\qty(\gamma_{\rm p} ; T_{\rm ej})$ is an integral given by
\begin{align} 
    \mathcal{I}\qty(\gamma_{\rm p} ; T_{\rm ej}) &=  \int_{r}^{\infty} \dd{x} \frac{x^2 - r^2}{e^x - 1}\, ,
    \\ &\simeq \begin{cases}
		\Gamma\qty(3) \zeta\qty(3) + r^2 \qty[\ln{\qty(1 - e^{-r})} - \frac{1}{2}] \ , & r \ll 1
		\\ 2\qty(1+r) e^{-r} \ , &r \gg 1
		\end{cases}
\end{align} 
where $r = \epsilon_{\Delta} / 2 \gamma_{\rm p} k_{\rm B} T_{\rm ej}$. \\

The hadronic interaction time with target nuclei $N$, $t_{\rm pN}$, is described by 
\begin{align} \label{eq:t_pp}
t_{\rm p N }\qty(\gamma_{\rm p}) = \qty[n_{\rm ej} \, \sigma_{\rm pN}\qty(\gamma_{\rm p}) \kappa_{\rm p N} c]^{-1} \, ,
\end{align}
where $n_{\rm ej}$ is the proton density of the ejecta, given by Eq.~\eqref{eq:density}, $\sigma_{\rm p N}\qty(\gamma_{\rm p})$ the proton-nucleus interaction cross section as function of the proton cosmic rays Lorentz factor. The interaction mean free paths are illustrated in Figure~\ref{fig:mfps}. 

Using Eq.\eqref{eq:t_pp}, we derive an analytical estimate of the hadronic interaction depth
\begin{align} \label{eq:pp_depth}
    \tau_{\rm p N}\qty(\gamma_{\rm p}) & = R_{\rm ej} \, n_{\rm ej} \, \sigma_{\rm p N}\qty(\gamma_{\rm p}) \, \kappa_{\rm pN} \, ,
    \\ &\sim 16 \  \qty(\frac{\beta_{\rm ej}}{0.3})^{-2} t_{3}^{-2} \qty(\frac{M_{\rm ej}}{10^{-4}M_{\odot}}) \left(\frac{\langle A\rangle}{100}\right)^{1/3}\left[\frac{\sigma_{\rm p p}\qty({\rm 1\,EeV})}{6.6 \times 10^{-26}{\rm \, cm}^{2}}\right] \ , \nonumber
\end{align}
 with the inelasticity $\kappa_{\rm pp} = 0.5$ . Here, we have used the superposition theory, assuming a dependency (which we also directly checked with EPOS) $\sigma_{\rm pN}\propto \langle A\rangle ^{2/3}$, for a mean mass number for nuclei in the ejecta $\langle A\rangle$. We have also used  Eq.~\eqref{eq:density}, with $n_{\rm ej}\propto 1/\langle A \rangle$. 
 
 Hence the hadronic interaction timescale only depends mildly on $\langle A\rangle$: $t_{\rm p N }\propto \langle A\rangle^{1/3}\sim 2$ for $\langle A\rangle=100$. 
One should caution that the outer layers of the ejecta could be dominantly composed of Hydrogen. As the relevant quantity here is the column density through the entire ejecta, the overall cross-section experienced by the particles would be lower. This effect could absorb the factor of $\langle A\rangle^{1/3}\sim 2$ quoted above.

All interaction mean free paths are presented at time $t=10^3\,$s in the top panels of Fig.~\ref{fig:mfps}.
The bottom panels of Fig.~\ref{fig:mfps}, illustrate the time evolution of the numerical mean free paths computed for proton and iron primaries with Lorentz factor $\gamma = 10^8$. One can see that at these energies, photopion production will operate efficiently from early times to several days and up to weeks. This is also valid at lower energies, until the photo-hadronic cross-section vanishes. The figure also shows that photopion production remains the dominant process over time for both proton and iron primaries. As a consequence the neutrino production is more likely to happen through photopion production (see section~\ref{section:neutrinos}) than hadronic processes. 

Note also that, as the proton-proton interaction mean free path is many orders of magnitude above the photo-hadronic interaction mean free path, adding or not a factor of $\langle A\rangle^{1/3}\sim 2$ would have no noticeable consequence. 
For simplicity, in the following, we will implement interaction cross-sections for Hydrogen backgrounds. For the interaction products, the superposition theory describes the interaction products between a projectile nucleus of mass and energy $(A_{\rm proj}, E_{\rm proj})$ and a target nucleus of mass $A_{\rm targ}$, as the same as for an interaction between a projectile $(A_{\rm proj},E_{\rm proj}/A_{\rm targ})$ and a target proton. This can result in a pile-up of lower-energy nucleons. The neutrinos produced by this process are however subdominant compared those produced by photo-hadronic interactions.

\subsection{Secondary pion and muon cascades} \label{subsection:meson_cascades}

\begin{figure}
\center
\includegraphics[width=0.49\linewidth]{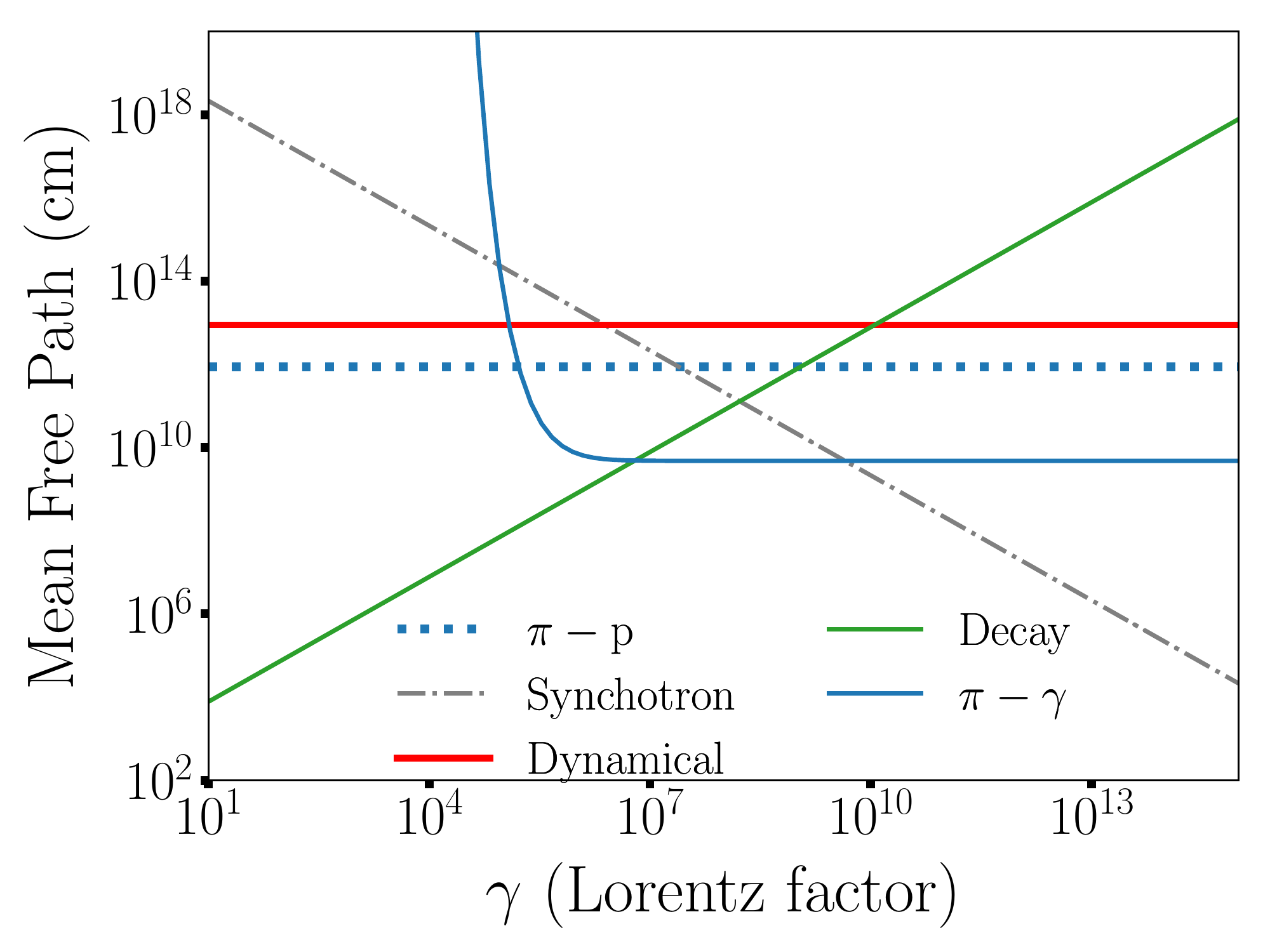}
\includegraphics[width=0.49\linewidth]{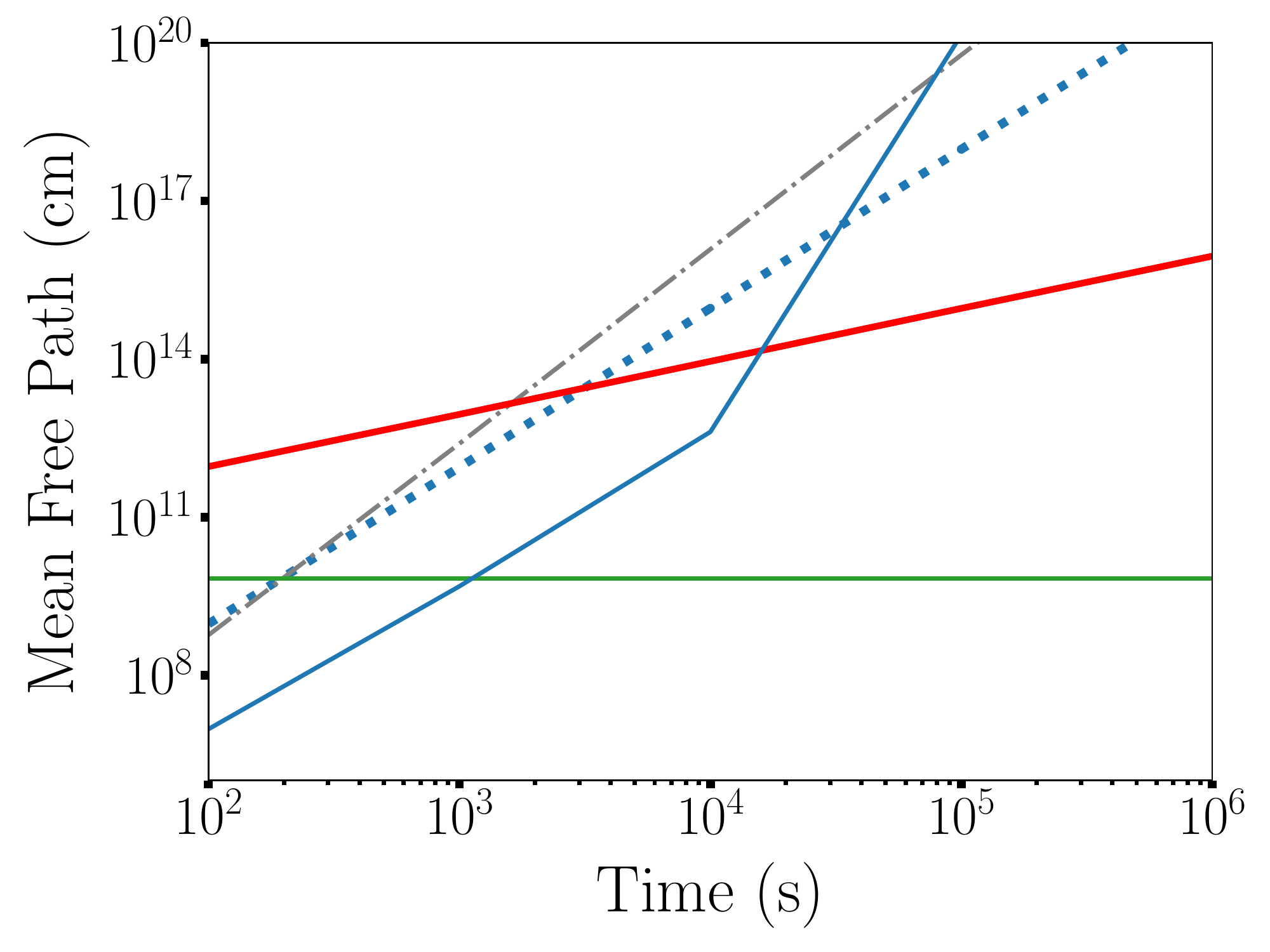}
\caption{Comparison between typical cooling/decay lengths for secondary mesons. {\it Left:} Typical cooling lengths for $\pi$ as a function of the particle Lorentz factor, at $10^{3}$\,s.  {\it Right:} Mean free paths for $\pi$ as function of time for a fixed Lorentz factor $\gamma = 10^7$.}
\label{fig:cs_mfp}
\end{figure}
Primary cosmic-ray interactions produce mesons through photo-nuclear and hadronic interaction channels (see section~\ref{subsection:cosmic_ray_interactions}). These mesons can in turn interact with radiative and hadronic backgrounds or undergo synchrotron radiation, before they decay and produce neutrinos. Meson cascades can therefore play an important role in the limitation of the neutrino flux. In the following, we focus on the charged pions cascades, as the pion channel is the most favored for neutrino production in $p-\gamma$ and $p-p$ interactions. In addition we include the description of muon cascades since their decay is responsible for $2/3$ of the neutrino production in the charged-pion decay chain.

Following the computations of section~\ref{subsection:cosmic_ray_interactions}, we calculate analytical estimates of energy-loss timescales of charged pions. The photon-mesons interaction time can be written as 
\begin{align}
t_{\pi \gamma}^{-1}\qty(\gamma_{\pi}) &= \frac{c}{2 \gamma_{\pi}^2} \int_{0}^{\infty} \dd{\epsilon_{r}} \epsilon_{r} \sigma_{\pi \gamma} \qty(\epsilon_{r}) K_{\pi \gamma} \qty(\epsilon_{r}) \int_{0}^{\epsilon_{r}/2 \gamma_{\pi}} \dd{\epsilon} \frac{n_{\rm photon}\qty(\epsilon)}{\epsilon^{2}} \nonumber \, ,
\\ &= 8\pi \, c \qty(\frac{k_{\rm B}T}{h \, c})^3 \qty( e^{- x_1}\qty(x_1 + 1) \expval{\sigma_{1} \, K_{1}} + e^{- x_2}\qty(x_2 + 1)  \expval{\sigma_{2} \, K_{2}} ) \, ,
\end{align}
where $x_1 = ( \epsilon_{a} - \epsilon_{b} ) / (2 \gamma \, k_{\rm B} T)$ and $x_2 = ( \epsilon_{b} - \epsilon_{c} ) / (2 \gamma \, k_{\rm B} T)$, with the following resonance parametrisation $\epsilon_{a}  = 0$, $\epsilon_{b} = 2 \, m_{\pi}c^2 $ and $\epsilon_{c} = 3 m_{\pi} c^2 $,  $\sigma_{1} = 10 \mu{\rm b}$,  $\sigma_{2} = 25 \mu{\rm b}$ \cite{Kaiser:2008ss} and assuming $K_1 = K_2 = 0.5$.

The pion-hadron and muon-hadron interaction times can be analytically estimated by
\begin{align}
t_{\rm \pi p}^{-1} &= n_{\rm p, ej} \, \sigma_{\pi p} \, K_{\rm \pi p}\, c  \ ,
\\ t_{\rm \mu p}^{-1} &= n_{\rm p, ej} \, \sigma_{\mu p} \, K_{\rm \mu p }\, c  \ ,
\end{align}
where $ \sigma_{\pi p} = 5 \times 10^{-26}$ cm$^2$,  $\sigma_{\mu p} = 1 \times 10^{-30}$ cm$^2$ and $K_ {\rm \pi p}= K_{\rm \mu p} = 0.8$ \cite{2018PhRvD..98d3020K}. For comparison, the typical decay times for pions and muons are $t_{\rm decay, \pi} \sim 2.6 \times 10^{-1}\,\gamma_{\pi, 7}\,{\rm s}$ and $t_{\rm decay, \mu} \sim 2.2\,\gamma_{\mu, 6}\,{\rm s}$.

Figure~\ref{fig:cs_mfp} shows the evolution of the typical lengths related to interactions, cooling processes and decay of charged pions in the optimistic scenario. The plot for muons is similar, with larger decay length and no photomeson interactions. On the left panel, we show the mean free paths as a function of Lorentz factor, for a given time $t=10^3$\,s after the merger. 
The lower intersection of the decay length with the typical length of any cooling/interaction process indicates the maximum energy at which charged pions or muons can decay and produce neutrinos.

In the left panel of Figure~\ref{fig:cs_mfp}, charged pions with Lorentz factor above $\gamma\sim 10^6$ do not directly decay but first undergo a cascade of photo-mesons processes (solid blue lines), losing energy and ending up in the left-hand side of this plot, where the decay length is the shortest.

The backgrounds evolve with time, leading to changes in the hierarchy of the interactions lengths. For a given Lorentz factor, charged pions and muons can produce neutrinos, only if they decay before losing energy or escaping the interaction region, i.e., only if their decay length is shorter than the typical lengths of the other processes. Hence, as illustrated in Fig.~\ref{fig:cs_mfp}, no direct decay is possible before $t=10^3$\,s for pions with $\gamma=10^7\,$, for the optimistic scenario.

For each Lorentz factor, there is a specific \emph{decoupling} time at which the decay processes become dominant. For each kilonova configuration parameter set (see Table~\ref{table:KN_parameters}) this decoupling time is different. The convolution of the decoupling time, corresponding to the highest secondary meson Lorentz factor, with the cosmic rays luminosity, $L_{\rm cr} = \eta_{\rm p} L_{\rm fb}$, where $\eta_{\rm p}$ is the baryon loading, the fraction of fallback luminosity dissipated to cosmic-ray luminosity, gives the maximum neutrino flux at the highest neutrino energy. In our case, since the cosmic-ray luminosity decreases as the fallback luminosity (with a time dependence of $t^{-5/3}$), kilonova configurations associated with the earliest decoupling times will lead to the highest neutrino flux scenarios.

\section{Neutrino fluxes from NS-NS mergers}\label{section:neutrinos}
In this section we derive the neutrino emissions from neutron stars merger events, taking into account all processes described in the previous section, as well as the remnant evolution in time. We first compute neutrino spectra for single sources and then integrate over the whole population of neutron star mergers to estimate the diffuse flux.


\subsection{Neutrino spectra from single sources}\label{subsection:neut_spec_single}
The neutrino emission from a single source can be analytically estimated using the interactions depths presented in Section~\ref{subsection:cosmic_ray_interactions} as
\begin{align}
E_{\nu}^{2} \dv{N_{\nu}}{E_{\nu}} = E_{\rm p}^{2} \dv{N_{\rm p}}{E_{\rm p}}\,f_\pi\,f_{\rm supp, \pi}
\left[\frac{1}{4} +\frac{1}{2}f_{\rm supp,\mu}\right]
\ ,
\end{align}
with ${\rm d}{N_{\rm p}}/{\rm d}{E_{\rm p}} = \mathcal{A} E_{\rm p}^{-\alpha} \exp\qty(- {E_{\rm p}}/{E_{\rm p, max}})$, the  injected cosmic-ray spectrum. $\mathcal{A}$ is a normalisation factor which scales with the cosmic-ray luminosity $L_{\rm cr}=\eta_{\rm p}L_{\rm fb}$, the duration of the emission $\Delta t$ and $E_{\rm min}$ and $E_{\rm max}$ the minimal and maximal injection energies: $\mathcal{A} = \qty(2 - \alpha) \times L_{\rm cr}\, \Delta t\, / \qty(E_{\rm max}^{2 - \alpha} - E_{\rm min}^{2 - \alpha})$. For numerical applications and for our results, we choose a value of $E_{\rm min}=10^{6}\,$GeV. $f_\pi = \min[1, \max(\tau_{p\gamma}, \tau_{pp})]\times 1/2$ or $2/3$ denotes the chance of the production of charged pions in  p$\gamma$ or pp interactions, respectively. $f_{\rm supp, *}$ describes the suppression of pions and muons due to cooling processes. 
Specifically, $f_{\rm supp, *} = 1 - \exp(-t_{\rm eff, *}/t_{\rm decay, *})$, $t_{\rm eff, *} = \qty(t_{\rm * \gamma}^{-1} + t_{\rm * p}^{-1} + t_{\rm *, sync}^{-1})^{-1}$ and $ t_{\rm decay, *} = \gamma_{\rm *} \tau_{\rm *}$, where $\tau_{*}$ is the mean life time of the particle $*$ and the subscript $*$ represents a charged pion or muon.

\begin{figure}[tbp]
\centering
\includegraphics[width=0.49\linewidth]{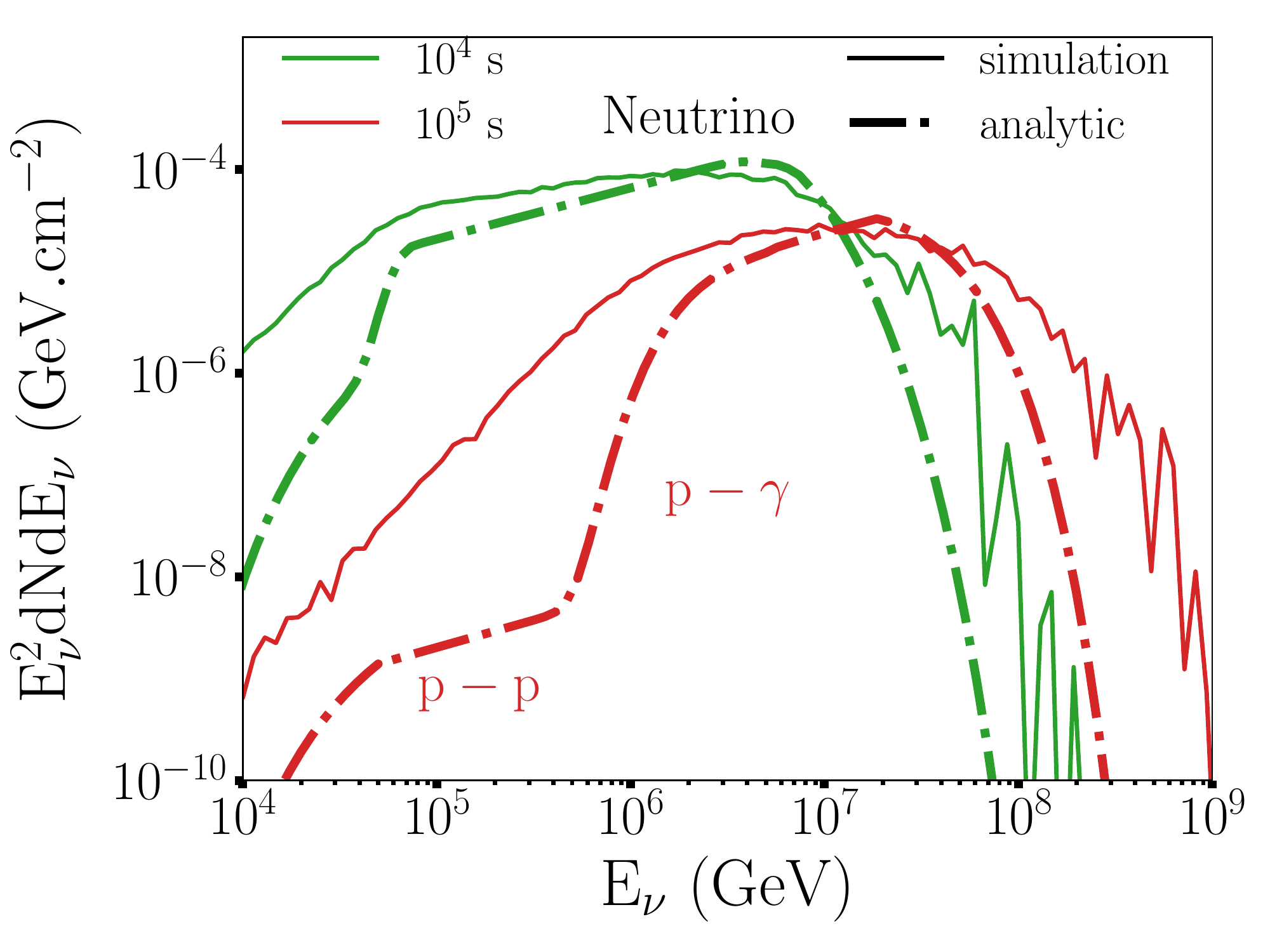}
\includegraphics[width=0.49\linewidth]{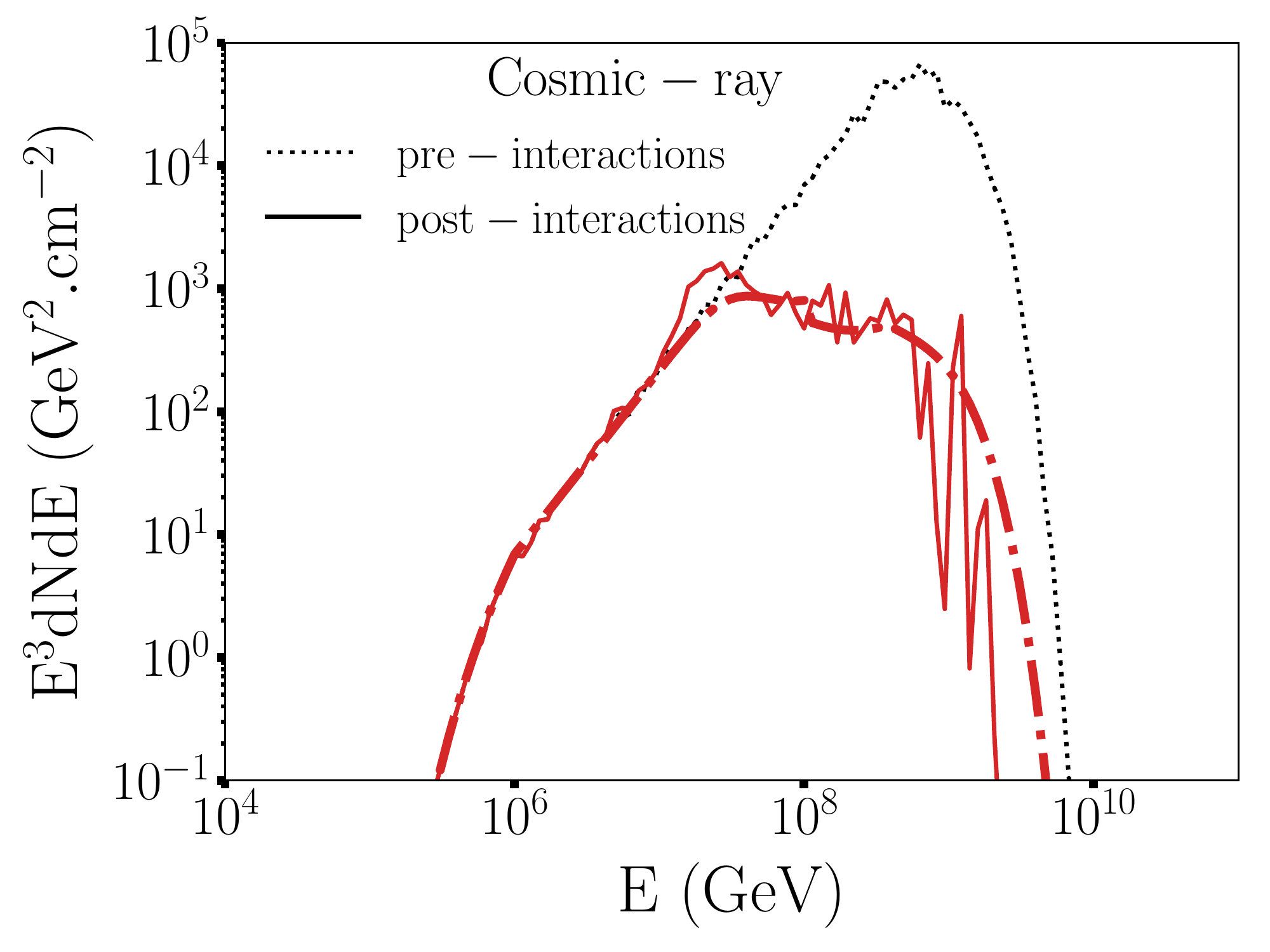}
\caption{Comparison between numerical simulations (solid) and analytical estimates (dot-dashed) for two different times $10^4$ (green) and $10^5\,$s (red) after the merger. {\it Left:} Neutrino spectra, {\it Right:} Cosmic-ray spectra. At $10^4\,$s, no cosmic ray survive the interactions. See text for details.}
\label{fig:numVSanalytic}
\end{figure}

Figure~\ref{fig:numVSanalytic} compares the numerical simulations and analytical estimates for two specific times after the merger: $t=10^4$\,s and $t=10^5$\,s. The right-hand panel represents the cosmic-ray spectra. The agreement between the numerical simulation and the analytical estimates is overall good, with errors below $1\%$. The left-hand panel represents the neutrino spectra. The mismatch between the numerical and analytical lines is larger and can be explained as follows.

First, the difference between the analytical estimates and the numerical simulations seen in the cosmic ray spectra at energies $\sim 10^{7}$\,GeV (for $t=10^5$\,s) is responsible for part of the mismatch around $E=10^6$\,GeV in the neutrino spectra (since about $5\%$ percent of the energy of the proton goes into neutrinos). In this energy range, the conversion of proton energy into neutrino energy is not well reproduced by the analytical estimate. The discrepancies at lower energies and in the high energy tail of the neutrino spectra is due to the photopion production model. Our analytical estimates only considers a constant interaction cross section for photopion production, while the accurate implementation of other channels smooth out the secondary particle energies over a wider range. Note however that the peak of each spectra is accurately reproduced, and  the good agreement in the cosmic-ray spectra implies that the fraction of proton energy converted into meson (pion) energy is correctly estimated. 

Regarding the cosmic rays spectra it can be seen that already at $t=10^5$\,s the primary cosmic rays undergo severe interactions leading to a large depletion of $\sim 2$ orders of magnitude between the pre and post interaction spectra.
At time $10^4\,$s, most cosmic rays lose energy via drastic photo-pion interactions, hence the absence of cosmic-ray flux at this time in the right-hand side plot. In the final picture, at early times ($>1$\,s) no cosmic rays can escape the kilonova as the number of interaction is too large, at longer times ($>10^4$\,s) a mixed composition appear and in between a transition from pure proton to mixed composition can be seen. However the diminution of the baryon loading with time result in a negligible cosmic-ray flux.

\begin{figure}[tbp]
\centering
\includegraphics[width=0.49\linewidth]{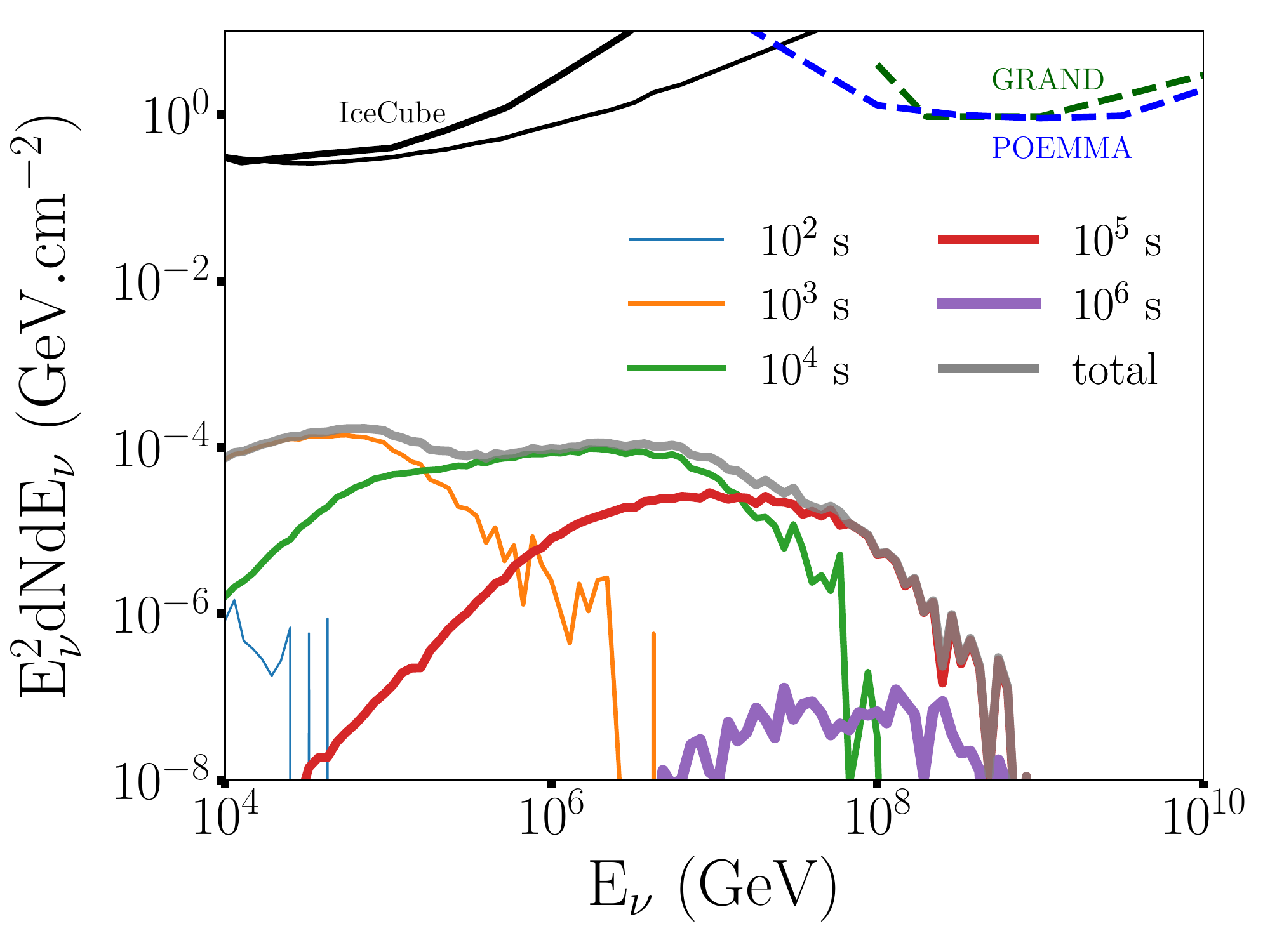}
\includegraphics[width=0.49\linewidth]{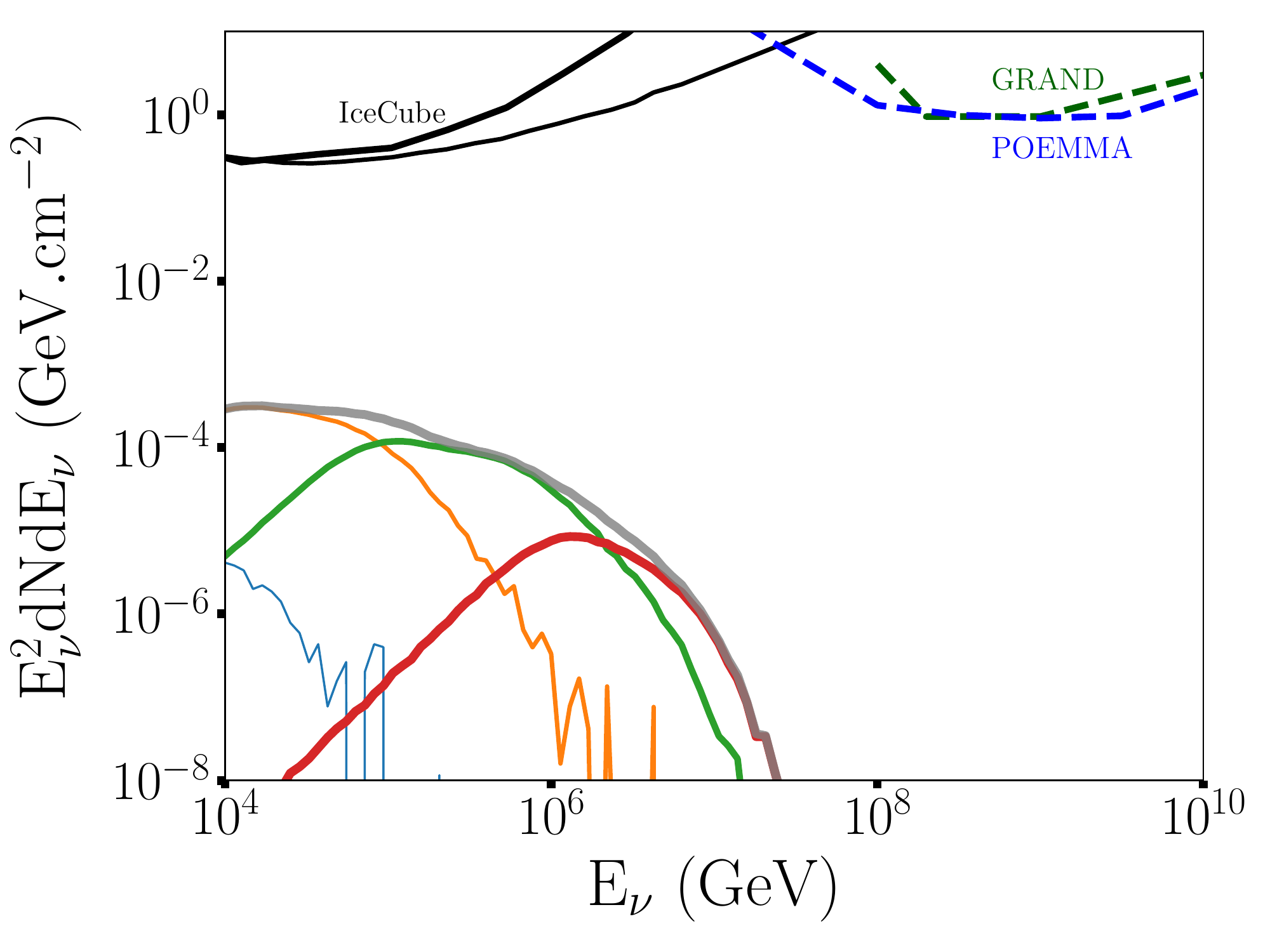}
\caption{Neutrino spectra for one source at distance $40$\,Mpc (optimistic scenario), for an injection spectral index $\alpha = 1.5$ and baryon loading $\eta_{\rm p} = 0.1$. Lines with increasing thickness represent neutrino fluences integrated up to the indicated (increasing) times after the merger. {\it Left:} pure proton injection and {\it Right:} pure iron injection. Black solid lines represent the IceCube point-source sensitivity for two declination configurations of the source in the sky: $0^\circ<\delta<30^\circ$ (best sensitivity) and $30^\circ<\delta<60^\circ$ \cite{Aartsen:2014gkd}. Dashed lines are projected point-source sensitivities for future experiments: POEMMA \cite{Venters:2019xwi} (blue) and GRAND (green) \cite{Alvarez-Muniz:2018bhp}.} 
\label{fig:nuflux_one_source}
\end{figure}

The numerical spectra for one source are shown in Figure~\ref{fig:nuflux_one_source} for different times after the merger, for pure proton (left) and pure iron (right) injections at the kilonova input, for the optimistic scenario. We can clearly identify an optimum neutrino production time around $t=10^3 - 10^4$\,s. The optimum time is the result of a  combination between i) a high cosmic-ray luminosity, ii) a high efficiency of cosmic-ray interactions leading to mesons production, and iii) a sufficiently low rate of meson (and muon) cascades leading to neutrino production. Consequently at earlier times, the neutrino flux is low (and with limited neutrino energy) because of the strong meson cascade rates and at later times it is low due to the decrease of primary interactions and also the decrease in the cosmic rays luminosity (scaling as the fallback luminosity $\sim t^{-5/3}$). Finally by comparing the left panel corresponding to proton primaries with the right panel corresponding to iron primaries, one can notice that in this model, the iron primaries leads to a higher neutrino flux at early times, with a softer spectral index overall. This can be explained by the accumulation at low energies of neutrinos produced by secondary nucleons created during the photonuclear interactions with the iron. Another noticeable effect is the lower maximal energies achieved by the neutrinos generated by iron primaries. This is due to fact that in our model, the maximal acceleration energy scales as $(A/Z)^{3/2}$ (Eq.~\ref{eq:Emax}), yielding roughly $E_{\rm Fe}^{\rm max} \approx 3 \times E_{\rm p}^{\rm max}$. However, the energy transferred to each pion through the photonuclear interactions scales as $E_{\rm Fe}^{\rm max}/A$. Consequently, the average pion energy will be lower by a factor $\sim 20$ for iron primaries compared to proton primaries, hence the lower maximal energies. Additionally, this difference implies that, at earlier times, iron-induced pions directly decay, while proton-induced pions cool via photo-hadronic interactions (see Fig.~\ref{fig:cs_mfp}). This effect also accounts for the slightly higher neutrino fluxes at low energies, at early times. Note that this effect no longer operates at times $t>10^4\,$s, because pions produced at the cosmic-ray maximal acceleration energy always directly decay (in other words, the photon background decreases sufficiently fast to compensate the increase of the maximal acceleration energy).
  
We overlay in these plots the IceCube fluence sensitivity, calculated from the effective area presented in \cite{Aartsen:2014gkd} for the optimal declination range $0^\circ<\delta < 30^\circ$ (thin lines), and for the declination range $30^\circ<\delta < 60^\circ$(thick lines). The IceCube-Gen2 effective area is projected to be $\sim 10^{2/3}$ times larger \cite{Aartsen:2014njl}. At ultra-high energies, we also indicate the projected sensitivities of the GRAND \cite{Alvarez-Muniz:2018bhp} and POEMMA~\cite{Venters:2019xwi} experiments.
The neutrino spectrum from single sources presents a plateau in the IceCube energy range ($10^{4-6}\,$GeV), at times $t\sim 10^{3-4}\,$s after merger. Even for optimistic scenarios, the low-flux levels would only allow detection with IceCube-Gen2 if the sources are located at distances $\lesssim 4\,$Mpc. 

\begin{figure}[tbp]
\centering
\includegraphics[width=0.49\linewidth]{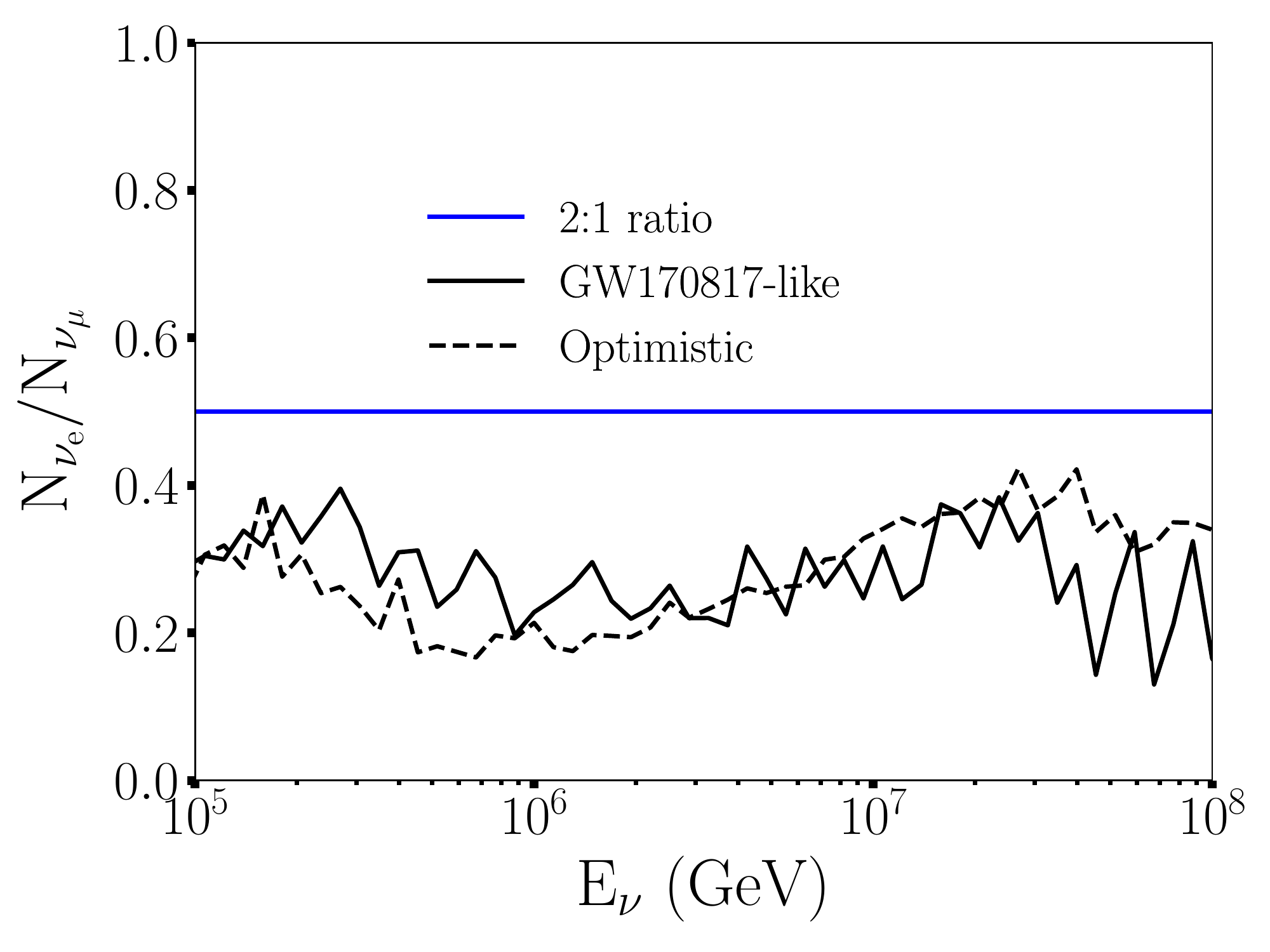}
\includegraphics[width=0.49\linewidth]{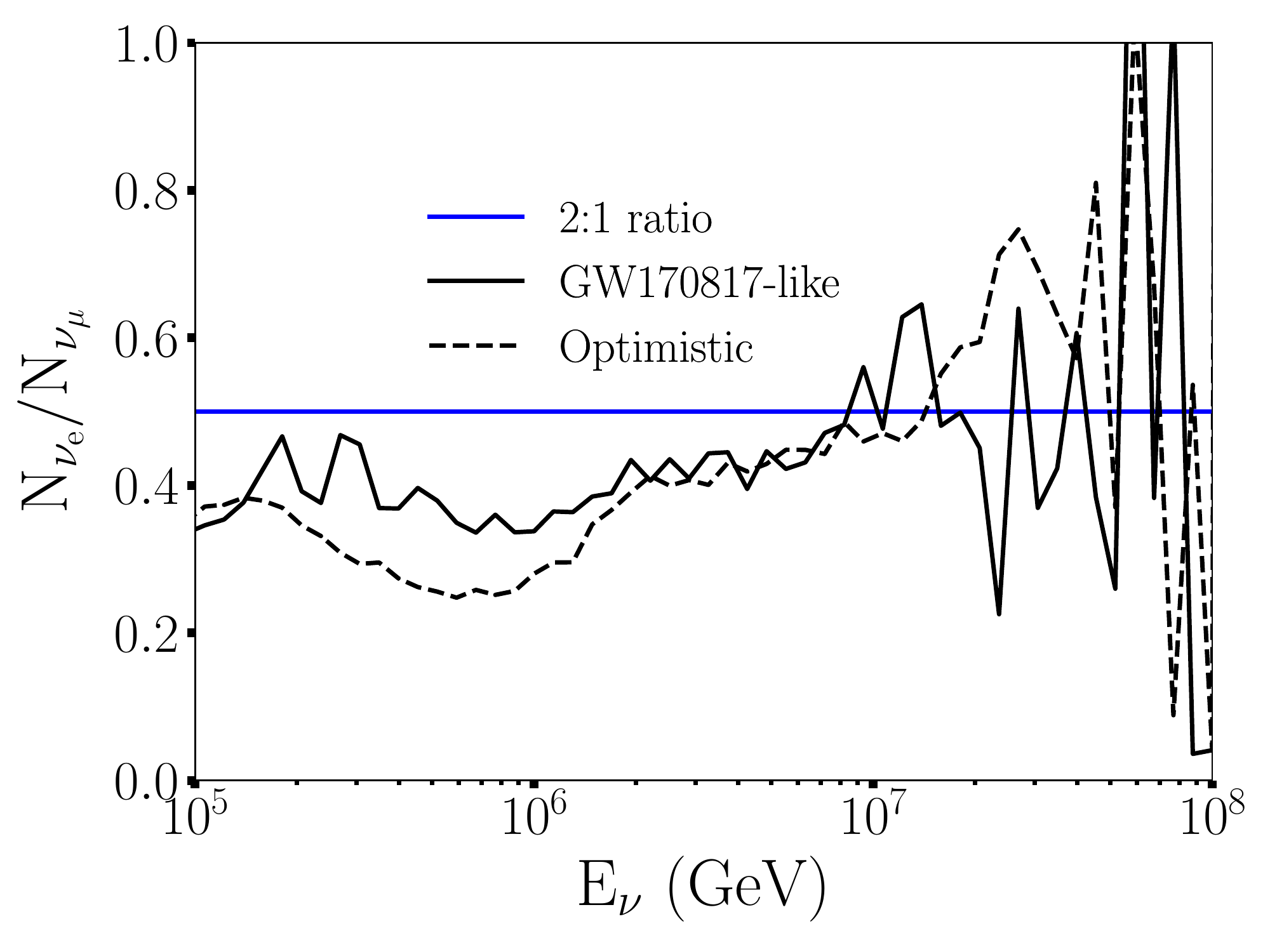}
\caption{Ratio of electronic neutrinos and muonic neutrinos as function of their energy for a pure proton injection ({\it left}) and a pure iron injection ({\it right}) at the source.}
\label{fig:flavor_ratio}
\end{figure}

Finally, Figure~\ref{fig:flavor_ratio} shows the ratio of electronic neutrinos and muonic neutrinos as function of their energy for a pure proton injection (left) and a pure iron injection (right) at the source (i.e at production). From \cite{Bustamente_2019}, the production of neutrinos through the decay of high-energy pions, which leads to a composition ratio of $1:2:0$ at the source, is favored by IceCube data. The scenario involving strong muon energy losses, which produces a ratio of $0:1:0$ at the source, is also slightly favored. As shown in figure~\ref{fig:flavor_ratio}, none of these two scenarios correspond to our models. Only the model with pure iron injection above $10^7$\,GeV approximately gives a ratio of $1:2:0$. However, we note that no neutrino detection has been performed at these energies and that this part of the spectrum presents a high statistical noise.
The deviation from expected scenarios of \cite{Bustamente_2019} roots in the high pion and muon energy losses (see section \ref{subsection:meson_cascades}) or in the mixed pion production channels involving both $p-\gamma$ and $p-p$ interactions (see section \ref{subsection:cosmic_ray_interactions}).

\subsection{Diffuse neutrino flux}

\begin{figure}[tbp]
\centering
\includegraphics[width=0.49\linewidth]{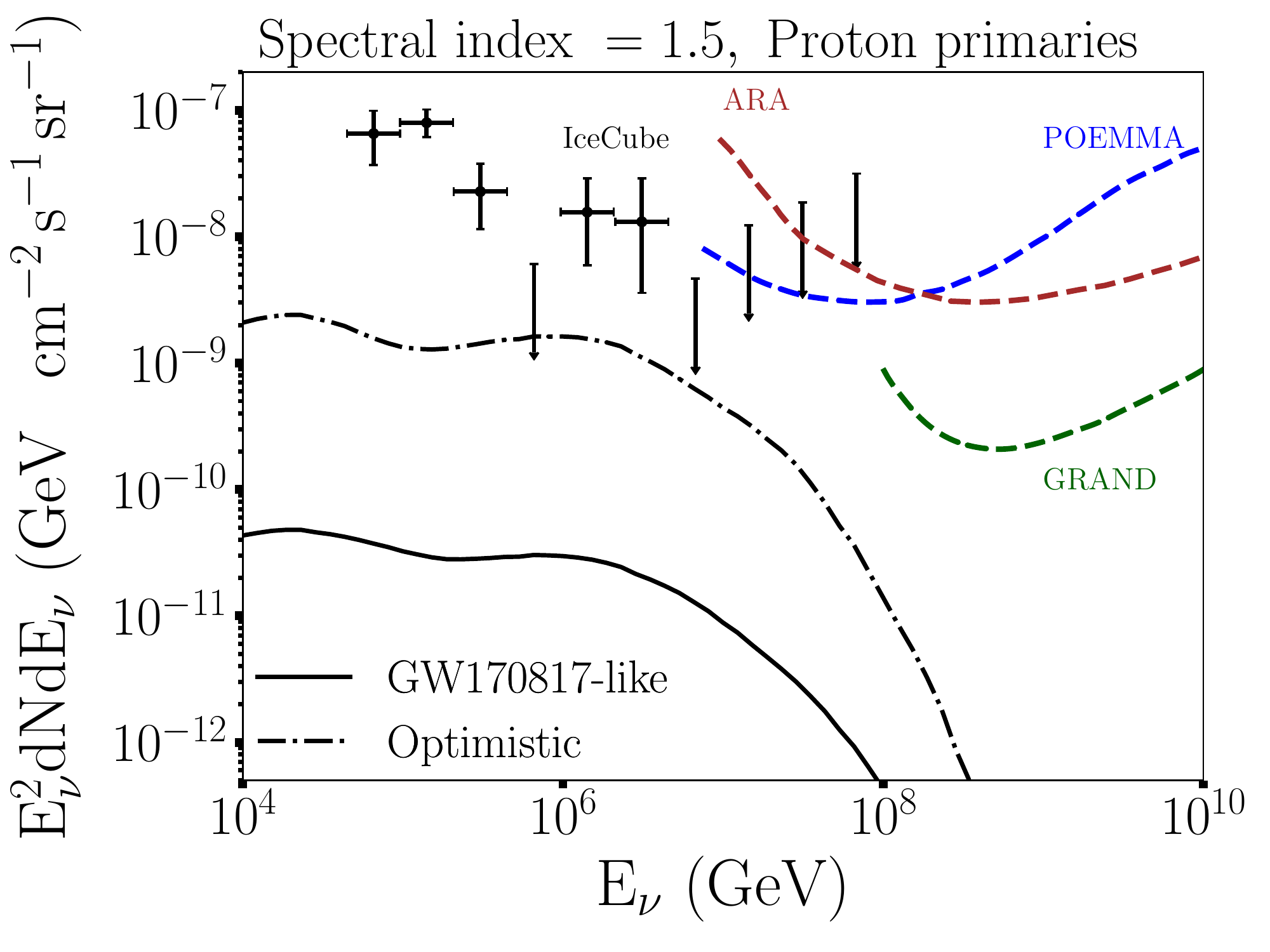}
\includegraphics[width=0.49\linewidth]{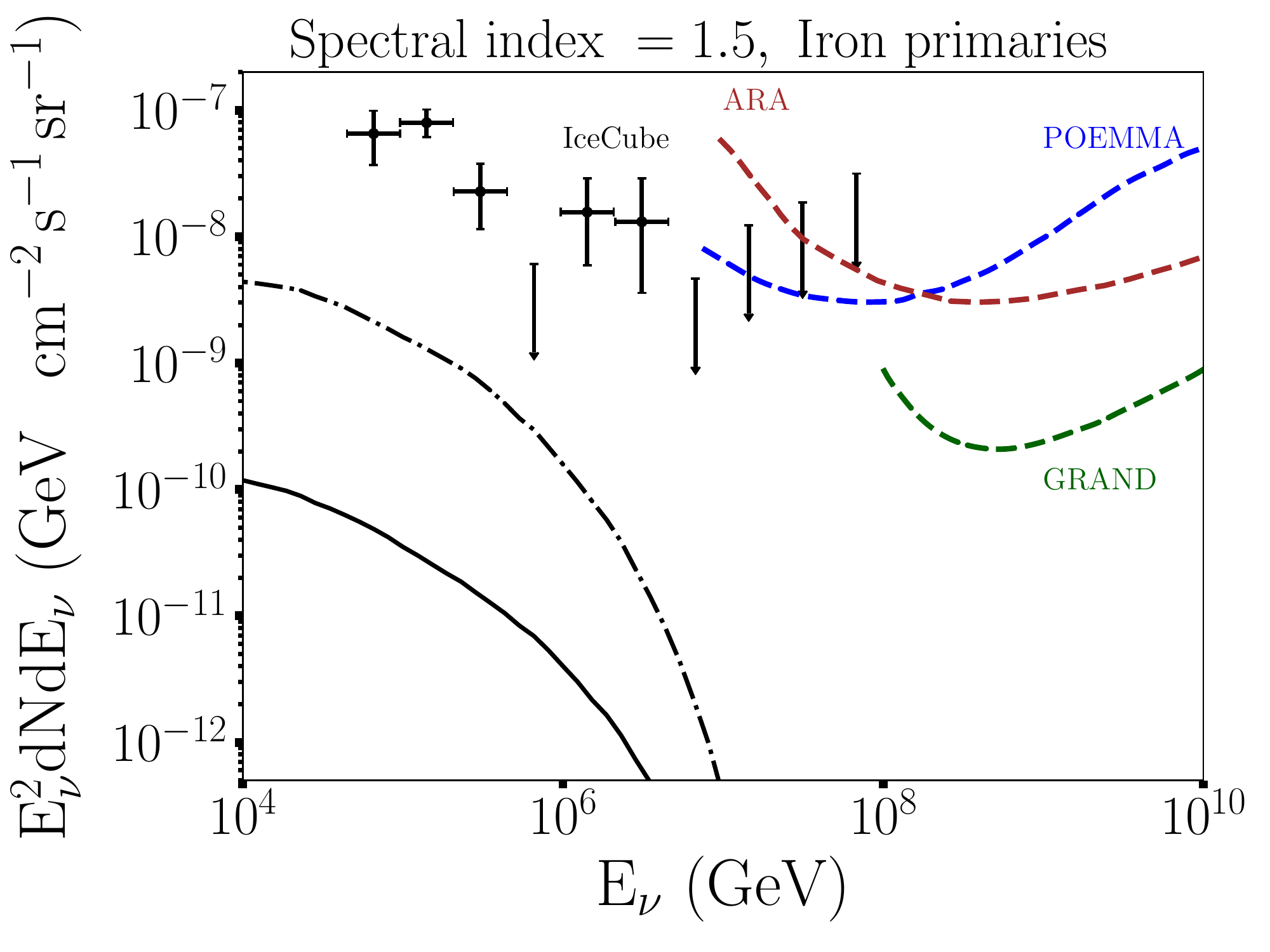}
\includegraphics[width=0.49\linewidth]{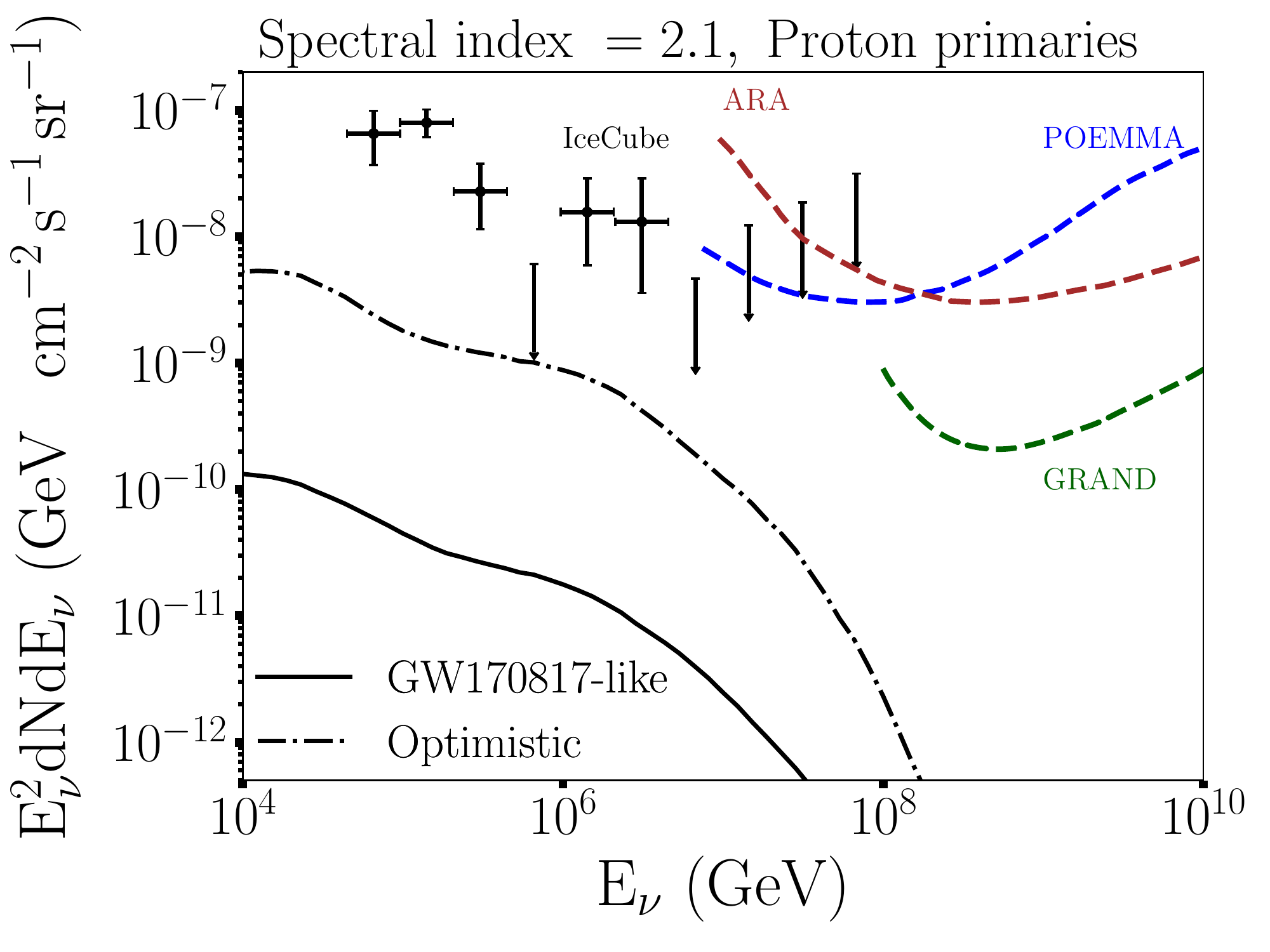}
\includegraphics[width=0.49\linewidth]{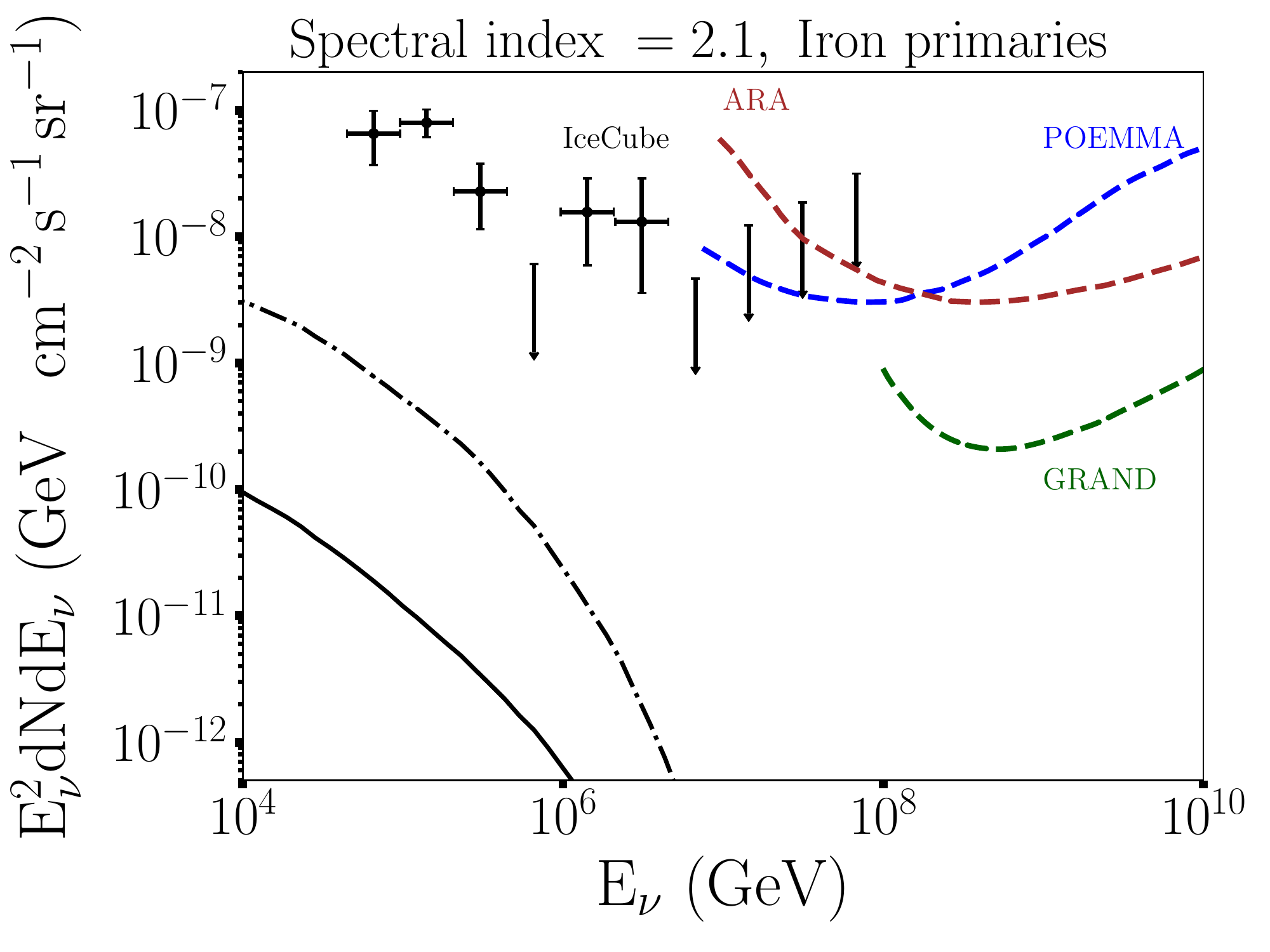}
\caption{Diffuse neutrinos spectra for injection spectral index $\alpha=1.5$ {\it (top)} and $\alpha=2.1$ {\it (bottom)} of proton primaries {\it (left)} and iron primaries {\it(right)}, with baryon loading $\eta_{\rm p} = 0.1$. The GW170817-like scenario follows a flat source evolution with rate $\dot{n}_0=600\,$Gpc$^{-3}$\,yr$^{-1}$, while the optimistic scenario follows a SFR source evolution with rate $\dot{n}_0=3000\,$Gpc$^{-3}$\,yr$^{-1}$.}
\label{fig:diffuse_nu_flux_spec15}
\end{figure}

Assuming that similar particle acceleration and interaction processes happen in  binary-neutron-star merger remnants, the integrated high-energy neutrino flux due to relativistic ions interacting with kilonovae photons is calculated by 
\begin{align}\label{eqn:diffnu}
\Phi \qty({E_\nu}) = \frac{c}{4 \pi H_{0}} \int_{0}^{\infty} \dd{z} \frac{\dot{n}\qty(z)}{\sqrt{\Omega_{\rm M} \qty(1+z)^3 + \Omega_{\Lambda}}}\frac{{\rm d}{N_\nu}}{{\rm d}{E'_\nu}}\qty[{E'_\nu} = (1+z){E_\nu}] \ ,
\end{align}
where ${\rm d}{N_\nu}/{\rm d}{E'_\nu}$ is the neutrino emission by one merger event at redshift $z$ in the energy ${E'_\nu}=(1+z) \,{E_\nu}$. In the above expression we have assumed a flat universe. We adopt the cosmological parameter values $\Omega_{\rm M}=0.315$, $\Omega_{\rm \Lambda}=0.685$, and the Hubble constant $H_0 = 67.4 \,{\rm km \,s}^{-1}\,{\rm Mpc}^{-1}$~\cite{Aghanim:2018eyx}. $\dot{n}(z)$ describes the rate of binary neutron-star mergers,  $\dot{n}\qty(z)= \dot{n}_{0} {\cal R}(z)$, with ${\cal R}(z)$ being the source emissivity evolution. 

We calculate the time-integrated neutrino flux from one source, ${\rm d}N_\nu/{\rm d}E_\nu$, by integrating over the neutrino fluences at time $t_i$ obtained numerically  (as in Figure \ref{fig:nuflux_one_source}): ${\rm d}N_\nu/{\rm d}E_\nu \approx \sum_i ({\rm d}N_\nu/{\rm d}E_\nu\,{\rm d}t_i)\,\Delta t_i$, with a resolution of $6$ logarithmic bins in time.

The local rate of NS-NS merger events was estimated to by LIGO/Virgo to lie in the range $\dot{n}_0\sim 110-3840\,{\rm Gpc}^{-3}\,{\rm yr}^{-1}$ (90\% confidence level), based on the single GW170817 event \cite{PhysRevLett.119.161101}. More refined estimates seem to converge towards rates of $\sim 600\,{\rm Gpc}^{-3}\,{\rm yr}^{-1}$ \cite{2017PhRvL.119p1101A, 2018MNRAS.474.2937C, 2018MNRAS.475..699S}. 

The cosmic evolution of NS-NS merger events is poorly known from observation. Recent theoretical studies based on neutron-star progenitors and binary-system evolution shows a flat behavior of the population with redshift (e.g., \cite{2014LRR....17....3P,2013ApJ...779...72D}). On the other hand, the connection of these systems with short GRBs, and the observation of the latter can indicate a source emissivity following the star formation rate (SFR) \cite{Ghirlanda:2016ijf,2018arXiv181109296C}. 

In order to be conservative, we adopt a flat evolution model with $\dot{n}_0=600$\,Gpc$^{-3}$yr$^{-1}$ for the GW170817-like scenario. Such a hypothesis represents the simplest model and do not presume of any enhancement of the population at earlier times in the universe history.  For the optimistic scenario, we assume a SFR evolution rate following Ref.~\cite{Hopkins:2006bw} and a local merger rate $\dot{n}_0=3000$\,Gpc$^{-3}$yr$^{-1}$. The SFR evolution can enhance the diffuse neutrino flux level by a factor of $\xi_z\sim2-4$ \cite{2010JCAP...10..013K}.

One can estimate the maximal diffuse neutrino flux expected in different energy ranges via 
\begin{align} \label{eq:diffuse_compute}
E_\nu^2\Phi_{\rm max}(E_{\nu}, \Delta t) &\sim \frac{c}{4 \pi H_{0}}\frac{3}{8}\xi_z \dot{n}_{0} \eta_{\rm p}L_{\rm fb} \Delta t 
\\ &\sim 1.9 \times 10^{-9} {\rm\ GeV\,cm}^{-2}\,{\rm s}^{-1}\,{\rm sr}^{-1}\,\eta_{\rm p, -1}\frac{\xi_{z}}{4} \\
&\qquad\qquad\times\qty(\frac{\dot{n}_{0}}{3000 {\rm\ Gpc}^{-3}{\rm yr}^{-1}}) \qty(\frac{L_{\rm cr}}{5.6 \times 10^{44} {\rm\ erg\,s}^{-1}})  \qty(\frac{\Delta t}{10^4\,{\rm s}}) \ , \nonumber
\end{align}
assuming a SFR evolution. At each time step, the neutrino flux peaks at one specific energy range and corresponds to a given cosmic-ray luminosity. The IceCube energy range $\sim E_\nu\gtrsim 10^{4}\,$GeV is reached from times $t=10^3\,$s, as can be seen in the time-dependent fluxes presented in Fig.~\ref{fig:nuflux_one_source}. 
In the optimistic scenario, assuming a $\eta_{\rm p}=10\%$ baryon loading, a SFR evolution with $\xi_z = 4$, the diffuse flux at $t=10^4\,$s estimated above corresponds to $\sim6\,\%$ percent of the observed IceCube flux of order $E_\nu^2\Phi_{\rm IC}\qty(E_{\nu}=10^6{\rm\,GeV}) \sim 3 \times 10^{-8}\,{\rm\ GeV\,cm}^{-2}\,{\rm s}^{-1}\,{\rm sr}^{-1}$.

Figure~\ref{fig:diffuse_nu_flux_spec15} presents the diffuse neutrino fluxes computed by equation~(\ref{eqn:diffnu}) for two representative cosmic-ray injection indices, $\alpha = 1.5$ and $\alpha= 2.1$ and two primary cosmic-ray compositions, proton and iron. In all panels, the dashed lines and solid lines correspond respectively to the optimistic scenario and the more conservative model based on the observation of GW170817. 

Consistent with the calculation above, we find that high-energy neutrinos from proton primaries can contribute to about $\sim 6\%$ of the diffuse flux measured by the IceCube Observatory~\cite{Aartsen_2016} in the optimistic scenario, for spectral index $\alpha =1.5$ and slightly lower for $\alpha=2.1$. Although the assumed event rate is 5 times higher than the standard value being recently advocated from the recent LIGO/Virgo data, the reasonable values of $\eta_{\rm p}$ leaves room for lower event rates, at the cost of higher baryon loading. In particular, in this optimistic model, a baryon loading of $20\%$ would enable to reach a larger fraction $12\%$ of the observed IceCube flux. With an order-of-magnitude better sensitivity, IceCube-Gen2 could be able to detect such underlying signals, in particular if cross-correlation analysis are done with other gravitational-wave and gamma-ray signals.

The iron primaries produce a slightly higher diffuse flux at low energies thanks to the accumulation of secondary nucleons, but the steeper spectra due to cut-offs at lower energies than for protons lead to overall lower diffusive fluxes. The neutrinos produced in the conservative GW170817-like scenario account for a negligible $<1\%$ fraction of the observed IceCube flux.

Muon neutrinos induce tracks in the detector and can accurately point to the sources, enabling stacked point-source searches. In that framework, one can calculate the muon-neutrino event rate of a  population of sources jointly observed in gravitational waves, as done in Ref.~\cite{Kimura18}. Following these authors, we estimate the expected number of muon neutrino events expected in IceCube-Gen2 by ${\cal N}_\mu=\int ({\rm d}N/{\rm d}E_\nu) A_{\rm eff}(\delta,E_\nu) {\rm d}E_\nu$, with $A_{\rm eff}(\delta,E_\nu)$ the instrument effective area. We consider only the upgoing and horizontal events that have declination $\delta>5^\circ$, as the detectability of downgoing events is uncertain for IceCube-Gen2. We assume that sources are standard candles with equal luminosity, and are distributed uniformly in the local Universe within $300\,$Mpc. Up to that distance, LIGO/Virgo should be sensitive enough to ensure the gravitational-wave detection of any occurring BNS merger as powerful as in our optimistic scenario (but also as in the GW170817-like model). In the optimistic scenario, the BNS event rate of $n_0=3000$\,Gpc$^{-3}$yr$^{-1}$ implies a local rate within 300\,Mpc of $\sim 340$ mergers per year. The neutrino flux in our model being isotropic, this yields a rate of $\sim 184$ observable events per year within the $\delta>-5^\circ$ sky. The expected rates in this optimistic configuration are of order $\sim 0.1\,{\rm yr}^{-1}$ with IceCube-Gen2, for baryon loading $\eta_{\rm p}=0.1$. One can thus expect a detection with several years of operation. 

The rates could be linearly scaled up with higher baryon loading. Also, taking into account downgoing events could double the rate, depending on the final experimental performances. Finally, as noted in Ref.~\cite{Kimura18}, it is likely that LIGO/Virgo significantly improves their sensitivity, increasing thereby the coincident detection rate.

\section{Conclusion, Discussion}
We have calculated the high-energy neutrino fluxes from binary neutron star merger remnants, focusing on their production by relativistic particles accelerated in unbound outflows from the central black hole accretion disk fed by the fall-back of marginally bound debris.  The relativistic particles generate neutrinos through their interaction with the kilonova radiation field and inner (red) kilonova baryon ejecta shell. We examined two scenarios.  The first scenario, motivated by observations and modeling of LIGO's first neutron star merger GW170817, adopts an ejecta mass of $M_{\rm ej} = 10^{-2}\,M_{\odot}$, a velocity of $\beta_{\rm ej} = 0.3$ and a mass accretion disk $M_{\rm fb} = 5 \times 10^{-2}\,M_{\odot}$. The second scenario assumes a much lower ejecta mass, along with a larger fall-back mass $M_{\rm ej} = 10^{-4}\,M_{\odot}$, $\beta_{\rm ej} = 0.3$ and  $M_{\rm fb} = 10^{-1}\,M_{\odot}$. The second scenario, called ``optimistic'', was designed to maximize the produced neutrino fluxes, within the parameter space of physically-allowed ejecta properties.  We use a Monte-Carlo code to model the propagation and interactions of accelerated particles inside the kilonova ejecta, thereby computing the time-dependant neutrino fluxes resulting from the configurations described above. 

The neutrino flux is found to peak in the IceCube energy range ($10^{4-6}\,$GeV), at times $t\sim 10^{3-4}\,$s after merger. Single sources have low fluxes that can only be detected with IceCube-Gen2 for optimistic scenarios, if located at distances $\sim 4\,$Mpc. The diffuse flux is below detectability for the GW170817-like scenario. For the optimistic scenario, we find that, for baryon loading $\eta_{\rm p}\sim 20\%$, the diffuse flux could contribute to $\sim 12\%$ of the observed IceCube flux. This leaves room for a future experiment with increased effective area such as IceCube-Gen2, to detect an underlying flux. In particular, cross-correlation searches with gravitational wave signals could lead to an enhanced signal-detection power below detection thresholds, using spatial and temporal information~\cite{Fang:2020rvq}. The detection of muon neutrinos from stacked local sources detected in gravitational waves is possible for several decades of operation, for the optimistic scenario with baryon loading $\eta_{\rm p}=0.1$, with event rates of 0.1 neutrinos per year with IceCube-Gen2 for population rates of $\dot{n}_0=3000$\,Gpc$^{-3}$yr$^{-1}$. As discussed in the previous section, these rates could become higher with higher $\eta_{\rm p}$, but also with experimental improvements: e.g., if downgoing events can be efficiently detected with IceCube-Gen2 and if LIGO/Virgo improves its detection sensitivity. 

We note that the actual uncertainties on the structure of the the inner region of the binary neutron star merger remnants still leave room for higher fallback luminosities than examined here, due to larger fallback ejecta masses or super-Eddington regime. In this case, the flux could be boosted of up to one order of magnitude and comfortably fit the observed IceCube diffuse flux. We caution that such fluxes would assume also the optimistic population rate and evolution scenario discussed in the previous section. Although such an exercise is speculative, it motivates the search for cross-correlation searches between neutrino, gravitational-wave and gamma-ray diffuse signals. It is also likely that more observations of single sources and of the overall population will soon put strong constraints on such scenarios. 

In our computation, no element heavier than iron was included. The matter falling back to the disk and thus ejected in the disk wind could ultimately be of heavy composition. Although the fraction of lanthanides is low (of the order of $X_{\rm r} = 10^{-3}-10^{-2}$) most of the mass becomes trapped in atomic numbers of $A\sim 100-200$. If acceleration happens in a region where the temperature is still $>10^9\,$K, and/or the entropy is particularly high, nuclei may have not time to assemble yet. In addition, the cross section for hadronic interactions of nuclei scales roughly as $A^{2/3}$. This implies that for such particles, the hadronic interaction timescales in the acceleration region would be shorter of more than an order of magnitude, i.e., $t_{A{\rm p}}\lesssim 0.1\,{\rm s}< t_{\rm acc}$ at $t=10^3\,$s (see numbers in Section~\ref{subsection:acc_outer}). These nuclei would hence be mostly disintegrated into nucleons and lower mass nuclei, which can then be accelerated to the energies calculated in Section~\ref{subsection:acc_outer}. Note that a surviving nucleus with large mass is not likely to have an important contribution to the neutrino flux calculated here. Indeed, the neutrino maximum energy is limited by a combination of the synchrotron cooling of the primary, scaling as $(A/Z)^{3/2}$ and the photonuclear interactions, scaling as $1/A$. So for heavy primaries, neutrinos could pile up due to numerous secondary nucleons, but at an overall energy roughly 2 orders of magnitude lower than for protons, where they would be dominated by atmospheric neutrinos.

In many existing models in the literature \cite{Kimura17, Kimura18, Biehl:2017qen, 2018PhRvD..98d3020K, Ahlers2019}, the production of high-energy neutrinos in neutron star mergers takes place in relativistic jets during the prompt or extended phases (times $t<10^3\,$s). 
In these models, the energetics of the emission are up to a couple of orders of magnitudes higher than in our model, as arises due to the earlier emission time (higher available luminosity) and the Lorentz boost. The escape of secondary mesons without catastrophic cascades is possible because of the more diluted radiation field at the larger distances where processes occur. However, the directional emission from the collimated relativistic jet could reduce the detection probability. In contrast, the emission in our model, where the deflection of the cosmic rays accelerated in the equatorial plane and crossing the kilonova, result in an isotropic emission and could be in principle connected to objects for which the prompt photon emission is not observed. This is the reason why the detection rates found in our optimistic case are of the same order as the ones computed in the works of Refs.~\cite{Kimura18, Ahlers2019}. 

Another interesting side product of this model could be gamma-ray photons, that could be produced through synchrotron and curvature radiations in the kilonova ejecta \cite{2013arXiv1305.0783P,Murase_2018}. At $10^3$~s when secondary neutrinos and gamma rays are produced, the fallback radiation rate is $L_{\rm fb} \sim 10^{46}\,\rm erg\,s^{-1}$. For a source located at $40\,\rm Mpc$, this corresponds to a secondary flux of $F_{\nu,\gamma} \sim 5.2\times10^{-8}\,\rm erg\,cm^{-2}\,s^{-1}$. This is below the point-source sensitivity of IceCube but close to that of {\it Fermi}-LAT. The observation of such gamma-ray emissions could probe the existence of non-thermal interactions inside the kilonova ejecta.

\acknowledgments
We are grateful to Enrico Barausse, Fr\'ed\'eric Daigne, Raphael Duque, Irina Dvorkin, Bruno Giacomazzo, Shigeo Kimura, Robert Mochkovitch, and Kohta Murase for very fruitful discussions during this study. This work was supported by the APACHE grant (ANR-16-CE31-0001) of the French Agence Nationale de la Recherche.  BDM acknowledges support from NASA (\#NNX16AB30G) and from the Simons Foundation through the Simons Fellows program (\#606260).

\bibliographystyle{JHEP}
\bibliography{NSM_bib}

\end{document}